\documentclass{amsart}

\usepackage{amsmath,amsfonts}
\usepackage{amssymb}
\usepackage{amscd,amsthm}
\usepackage{graphicx}

\theoremstyle{plain} 
\newtheorem{theorem}{Theorem}[section]
\newtheorem{corollary}[theorem]{Corollary}
\newtheorem{lemma}[theorem]{Lemma}
\newtheorem{proposition}[theorem]{Proposition}

\newtheorem{remark}[theorem]{Remark}

\newtheorem{conjecture}[]{Conjecture}
  
\numberwithin{equation}{section}
\newcommand{\epscrit}[1][\omega]{\mathop{\epsilon_{crit} \left( #1
\right)}}
\newcommand{\rcrit}[1][\omega]{\mathop{\rho_{crit} \left( #1 \right)}}
\newcommand{\Brf}[1][\omega]{\mathop{{\it B}\left( #1 \right)}}
\newcommand{\R}{\mathbb{R}}
\newcommand{\C}{\mathbb{C}}
\newcommand{\Z}{\mathbb{Z}}
\newcommand{\N}{\mathbb{N}}
\newcommand{\T}{\mathbb{T}}
\newcommand{\Q}{\mathbb{Q}}
\newcommand{\D}{\mathcal{D}}
\newcommand{\normm}[2]{\lvert\lvert #1 \rvert\rvert_{#2}}

\renewcommand{\L}{\mathcal{L}}
\newcommand{\bigo}[1]{\mathop{\mathcal{O}\left( #1 \right)}}

\begin{document}

\vspace{-30em}
SUBMITTED TO: NONLINEARITY
\vspace{+3em}

\title[Scaling law in the Standard Map critical function.]{Scaling law in the Standard Map critical
function. Interpolating Hamiltonian and frequency map analysis.}

\author{Timoteo Carletti, Jacques Laskar}

\date{\today}

\address[Timoteo Carletti]{Dipartimento di Matematica ``Ulisse Dini'',  Viale Morgagni 67/A, 50134
Florence, Italy \\
and 
Astronomie et Syst\`emes Dynamiques, IMC, CNRS EP1825,
$77$ Avenue Denfert Rochereau, $75014$ Paris, France}

\address[Jacques Laskar]{Astronomie et Syst\`emes Dynamiques, IMC, CNRS EP1825,
$77$ Avenue Denfert Rochereau, $75014$ Paris, France}

\email[Timoteo Carletti]{carletti@math.unifi.it}
\email[Jacques Laskar]{laskar@bdl.fr}



\begin{abstract}
We study the behaviour of the Standard map critical function in a neighbourhood of a fixed
resonance, that is the scaling law at the fixed resonance.

\indent
We prove that for the fundamental resonance the scaling law is linear. We show numerical evidence
that for the other resonances $p/q$, $q \geq 2$, $p \neq 0$ and $p$ and $q$ relatively prime, the
scaling law follows a power--law with exponent $1/q$.
\end{abstract}

\maketitle

\section{Introduction.}
\label{introd}

The standard map \cite{Chirikov1,Greene1} is the area--preserving twist map of the cylinder:
$\R~/~(2\pi\Z)~\times~\R$ given by:
\begin{equation}
\binom{x^{\prime}}{y^{\prime}}= T_{\epsilon}\binom{x}{y}=\binom{x+y+\epsilon \sin x}{y+\epsilon \sin
x} 
\label{standardmap}
\end{equation}
where $\epsilon$ is a real parameter. Note that for all values of $\epsilon \in \R$ one has $S\circ
T_{\epsilon}\circ S=T_{-\epsilon}$ where $S$ is the involution of the cylinder
$S\binom{x}{y}=\binom{x+\pi}{y}$. This allows us to restrict the study of $T_{\epsilon}$ to
$\epsilon$ non negative.

\indent
The dynamics of the Standard Map depends critically on the parameter $\epsilon$. For $\epsilon=0$
the curves $y_0= constant$ are preserved by the map and the motion on these curves is a rotation of
frequency $\omega = y_0$. For $\epsilon >0$ not all these curves persist. Given an initial condition
$\left( x_0,y_0 \right)$ we define the {\em rotation number} $\omega_T$ by: 
\begin{equation*}
2 \pi \omega_T \left( x_0,y_0,\epsilon \right)=\lim_{n \rightarrow \infty}\frac{\pi_1\left(
T_{\epsilon}^n \left( x_0,y_0\right) \right) -x}{n}
\label{eq:rotnumbdef}
\end{equation*}
when the limit exists, where $\pi_1$ is the projection on the first coordinate: $\pi_1\left( x,y
\right)=x$. For $\epsilon >0$ by Poincar\'e--Birkhoff's Theorem (see
~\cite{poincare2,birkoff1,birkoff2,herman1} and \cite{Mather2} for a different proof) homotopically
non--trivial invariant curves with rational rotation number do not exist: the invariant curves of
rotation number $\omega_T \left( x,y,0\right)=p/q$, for integer $p,q$, $q>0$ relatively prime, split
into $2q$ points when $\epsilon >0$. This set of points is divided into two subsets of points each
one composed of $q$ points: the {\em elliptic} points and the {\em hyperbolic} ones.

\indent
Let us define
\begin{align*}
\epscrit=\sup \{&\epsilon>0:\forall \epsilon^{\prime}\in \left[ 0,\epsilon \right], \text{there
exists a $\mathcal{C}^1$, homotopically}\notag \\ &\text{ non--trivial, invariant circle
of $T_{\epsilon^{\prime}}$ whose rotation number is $\omega$} \}
\label{def:epscrit}
\end{align*}
and $\epsilon_{crit}\left(\omega\right)=0$ if $\omega \in \mathbb{Q}$. The function $\omega \mapsto
\epscrit$ is called the {\em critical function} for the standard map. The definition of $\epscrit$
actually depends on the smoothness required for the invariant curve. However, whatever is the
regularity of the curve (say $\mathcal{C}^1$, $\mathcal{C}^k$, $\mathcal{C}^{\infty}$ or
$\mathcal{C}^{\omega}$) this function is very irregular: it is almost everywhere discontinuous and
zero on a dense $G_{\delta}$--set. Moreover J.Mather proved~\cite{Mather} that $\sup_{\omega \in
\R}\epscrit \leq \frac 4 3$ and it is conjectured that $\sup_{\omega \in \R}\epscrit =
\epsilon_{crit}\left( \frac{\sqrt{5}-1}{2}\right)=0.971635\ldots$~\cite{Greene1,LaskarFroeschleCelletti,Laskar4}.

\indent
For suitable irrational $\omega$, KAM theory~\cite{Kolmogorov,Arnold,Moser} assures that for
sufficiently small $\lvert \epsilon \rvert$ there exists an analytic invariant circle upon which the
dynamics of $T_{\epsilon}$, restricted on this curve, is analytically conjugated to the rotation
$R_{\omega}\left( \theta \right)= \theta +2 \pi \omega$. This means that there exists an analytic
(w.r.t. $\theta$ on $\T=\R / 2\pi \Z$ and $\lvert \epsilon \rvert < \epsilon_0$, for some
$\epsilon_0 >0$) function $u \left( \theta, \omega, \epsilon \right)$ such that
\begin{equation*}
x = \theta + u \left( \theta, \omega, \epsilon \right),\quad 1+\frac{\partial u}{\partial
\theta}\left( \theta,\omega,\epsilon\right) >0 \quad \forall \theta \in \T
\end{equation*}
and $\theta^{\prime}=\theta + 2\pi \omega$, namely $u$ reparametrizes the invariant curve. Using
perturbation theory we can expand this function into a  power series w.r.t. $\epsilon$  and Fourier' series w.r.t. $\theta$
\begin{equation}
u \left( \theta, \omega, \epsilon \right)=\sum_{n \geq 1}\epsilon^n u_n\left( \theta, \omega
\right)=\sum_{n \geq 1}\epsilon^n \sum_{k \in \Z,\lvert k \rvert \leq n} \hat{u}_n^{(k)}\left(
\omega \right) e^{i k \theta}
\label{Lindstedt}
\end{equation}
 which gives the so called {\em Lindstedt series} \cite{poincare1}. One may define the {\em critical
radius of convergence} of such series as
\begin{equation}
\rcrit=\left[ \sup_{\theta \in \T} \limsup_{n \rightarrow \infty} \lvert u_n\left( \theta, \omega
\right) \rvert^{\frac{1}{n}} \right]^{-1}.
\label{eq:criticalradius}
\end{equation}
It is clear that $\rcrit \leq \epscrit$, but it is an interesting open question to decide whether
the two functions are equal or not on some set of frequencies. See \cite{berrettichierchia1,BCCF}
for a positive numerical answer and \cite{Davie2} for a negative one. This will imply that the
function defined for $\epsilon \in \mathbb{D}_{\rcrit}$ as the sum of the series ~\eqref{Lindstedt}
can be analytically continued outside its disk of convergence.

\indent
In \cite{MarmiStark} Marmi and Stark studied the relation between the $\epscrit$ and a ``universal
function'' depending only on the frequency: the Brjuno function $\Brf$ (see
\cite{Brjuno,MMY,yoccoz2}). Using the Greene residue criterion (\cite{Greene1}), Marmi and Stark
find {\em numerically} evidence that it exists $\beta >0$ and a positive constant $C$, such that
\begin{equation}
\Big\lvert \ln \epscrit + \beta \Brf \Big\rvert \leq C
\label{eq:marmistark}
\end{equation}
uniformly in $\omega$, as $\omega$ ranges in some set of quadratic irrationals, with $\beta$
approximately equal $0.9$. Comparing their results with similar problems in \cite{Marmi1,yoccoz2}
they expected the value $\beta=2$.

\indent
In \cite{BerrettiMarmi} and more recently in \cite{BerrettiGentile1} the behaviour of $\rcrit$ when
$\omega$ goes to a rational value is studied; in both cases {\em complex frequencies} were used. It
is well known that if the imaginary part of $\omega$ is not zero the series (\ref{Lindstedt})
converges because there are no small divisors. In these papers the limit $\omega \rightarrow \frac p
q$ of $\rcrit$ when $\omega = \frac p q + i\eta$ for some real $\eta \rightarrow 0$ was studied. The
result is the following scaling law
\begin{equation}
\rcrit \sim C_{p/q}\Big\lvert \omega - \frac p q \Big\rvert ^{\frac 2 q}
\label{eq:berrettigentilescalinglaw}
\end{equation}
for some positive constant $C_{p/ q}$ depending on the resonance.

\indent
Finally very recently Berretti and Gentile \cite{BerrettiGentile2} proved that
\begin{equation}
\label{eq:berrettigentile2}
\Big\lvert \ln \rcrit + 2 \Brf \Big\rvert \leq C^{\prime}
\end{equation}
for some positive constant $C^{\prime}$, uniformly w.r.t. $\omega \in \R$. This implies that the
series given in (\ref{Lindstedt}) converges if and only if $\omega$ is a Brjuno number (i.e. $\Brf <
\infty$).

\indent
The scaling law is related to this result which makes a link between rotation number and Brjuno
function. One can prove (~\cite{carletti}) that 
\begin{equation*}
  \liminf_{\substack{\omega\rightarrow p/q\\ \omega \text{ is a Brjuno's number}}} \left(\Brf+\frac
1 q \ln \Big\lvert \omega-\frac p q \Big\rvert\right)=c^{\prime}_{p/q}
\end{equation*}
with $c^{\prime}_{p/q}$ some non negative finite constant depending on the involved
resonance\footnote{\label{ftn:scallawbrjuno}We point out that the previous result is false if we
replace $\liminf$ with $\lim$. In fact considering the fundamental resonance for simplicity and
taking the sequence of Brjuno's numbers $\left( \omega_n\right)_{n\geq 0}$,
$\omega^{-1}_n=a_1\left(n\right)+\frac{1}{a_2\left(n\right)+\xi}$, with $a_i:\mathbb{N}\rightarrow
\mathbb{N}^*$ strictly monotone increasing functions for $i=1,2$ and $\xi$ any Brjuno's number. Then
we have $\lim_{n\rightarrow +\infty}B\left(\omega_n\right)+\ln \omega_n=0$ if $\ln
a_2\left(n\right)={\it o}\left(a_1\left(n\right)\right)$ for $n\rightarrow +\infty$, whereas if
$a_1\left(n\right)={\it o}\left(\ln a_2\left(n\right)\right)$ for $n\rightarrow +\infty$, then
$\lim_{n\rightarrow +\infty}B\left(\omega_n\right)+\ln \omega_n=+\infty$.}, then the result of
Berretti and Gentile \cite{BerrettiGentile2} implies the value $\beta =2$.

\indent
To end this short review about standard map critical functions and their scaling laws we cite the
very recent work of Treschev and Zubelevich \cite{TreschevZubelevich}, where they  studied (quite
general) area preserving twist maps of the standard cylinder into itself, $\epsilon$--close to an
integrable ones. They proved that under some assumptions: given any rational number $p/q \in \Q$,
there exist two positive constants $c$ and $\epsilon_0$ such that for all $\lvert \epsilon \rvert <
\epsilon_0$, there exist two invariant homotopically non--trivial invariant $\mathcal{C}^1$ curves
with frequencies $\omega_{l}$ and $\omega_{b}$ and
\begin{equation*}
 \omega_{l} < \frac p q < \omega_{b} \quad \text{and} \quad \lvert \omega_{b} - \omega_{l}\rvert < c
\epsilon.
\end{equation*}

\indent
The Theorem they prove (Theorem $3$ p.$76$ \cite{TreschevZubelevich}) is based on a strong
hypothesis (in particular assumption $2.$, which depends on the involved resonance) which is not
verified by the standard map, except for the resonance $0/1$.

\indent
We consider the fundamental resonance $0/1$, we fix $\epsilon >0$ and we define
\begin{align}
\omega^+\left( \epsilon \right)=\inf \{ \omega >0: T_{\epsilon}\text{ has a $\mathcal{C}^1$, homotopically non--trivial,} \notag \\ \text{ invariant curve of rotation number
$\omega$} \}
\label{def:omegap}
\end{align}
and 
\begin{align}
\omega^-\left( \epsilon \right)=\sup \{ \omega <0: T_{\epsilon}\text{ has a $\mathcal{C}^1$, homotopically non--trivial,} \notag \\ \text{invariant curve of rotation number
$\omega$} \}
\label{def:omegam}
\end{align}
note that $\omega^+\left( \epsilon \right)=-\omega^-\left( \epsilon \right)$. 
These definitions are similar to the one of $\mu(\omega)$ given by R.S. Mac Kay~\cite{mackay} in the study of the Hamiltonian of Escande.
The existence of a
stochastic layer around the hyperbolic fixed point $x=y=0$, implies $\omega^+\left( \epsilon
\right)\neq 0$, namely there is a ``gap'' in the curve which associates to each invariant circle its
rotation number (frequency curve). The main result of the first part of this paper is the following
Theorem\footnote{We point out that already in~\cite{mackay} for a one degree of freedom Hamiltonian system depending periodically on time, logarithmic singularities were found for the breakup of invariant tori with frequencies closer and closer to rational ones.}.
\begin{theorem}
\label{the:maintheorem}
There exist two positive constant $\Omega_{0/1}$ and $c_{0/1}$ such that for all $\lvert \omega
\rvert < \Omega_{0/1}$ then
\begin{equation*}
\Big\lvert \ln\epscrit - \ln \lvert \omega\rvert \Big \rvert \leq c_{0/1}.
\end{equation*}
\end{theorem}

Following~\cite{TreschevZubelevich}, to prove our Theorem we need a upper bound for $\ln \epscrit
-\ln \lvert\omega\rvert$. We obtain this bound combining ideas taken from two different domains: the exponential
smallness of the splitting of the separatrices and the interpolation of diffeomorphisms by vector
fields.

\indent
The splitting of the separatrices for the standard map has been widely studied. Using the results of
Lazutkin and Gelfreich~\cite{Gelfreich,Lazutkin2,Lazutkin,Lazutkinschtab} we get a lower bound to
the action corresponding to the last invariant torus. More precisely for a fixed $\epsilon$ we
define $y^+\left( \epsilon\right)$, respectively $y^-\left( \epsilon\right)$, as the intersection
point of the homotopically invariant $\mathcal{C}^1$ curve of frequency $\omega^+\left(
\epsilon\right)$, respectively $\omega^-\left( \epsilon\right)$, with the $y$--axis, note that
$y^+\left( \epsilon\right)=-y^-\left( \epsilon\right)$. Then we prove
(Proposition~\ref{pro:initialdata}) that
\begin{equation*}
y^+\left( \epsilon\right) \geq ce^{-\frac{\pi^2}{2\sqrt{\epsilon}}}
\end{equation*}
for some positive constant $c$ and $\epsilon$ small enough.

\indent
Using a result due to Benettin and Giorgilli~\cite{BenettinGiorgilli}, we construct an Hamiltonian
system whose time--$1$ flow interpolates the Standard map; this allows us to exhibit an asymptotic
development for the Standard map rotation number (Proposition~\ref{prop:asymptoticexp}) which, joint
with the bound on $y^+\left( \epsilon \right)$, leads to the desired upper bound. We note that to
prove our result it will be sufficient to find an interpolating Hamiltonian system $\epsilon^{\frac
3 2}$--close to the standard map; we nevertheless give the asymptotic development because to the
best of our knowledge this result was not know and, secondly, the construction of an interpolating
vector field exponentially close needs the same amount of difficulty as the construction of an
$\epsilon^k$--close vector field, for any power $k$.

\indent
In the second part (section~\ref{secnaff}) we present the {\em Frequency Map Analysis} (FMA) method
of Laskar \cite{DumasLaskar,Laskar2,Laskar1,LaskarFroeschleCelletti} and its application to the
numerical investigation of the Standard map critical function.

\indent
For the fundamental resonance $0/1$, we find numerical results in agreement with the analytical ones
presented in section~\ref{sec:main}. For the other resonances we find the behaviour
\begin{equation*}
\epscrit \sim c_{p /q}\Big\lvert \omega - \frac p q \Big\rvert ^{\frac 1 q}
\label{1law}
\end{equation*}
for $\omega$ in a neighbourhood of $p/q$ and $c_{p /q}>0$. We remark that using FMA we deal with
{\em real frequencies}.

\indent
Using the FMA tool we can do investigations directly on the critical function and go deeply in the
resonant region.  Actually we deal with frequencies that are only $6 \cdot 10^{-4}$ far from the
resonant value (in \cite{MarmiStark} the use of the Greene's residue criterion gives frequencies
that are $10^{-2}$ close to the resonant value). The result we show in this paper for the critical
function is in agreement with the one of \cite{MarmiStark}.

\indent
Supported by considerations on both methods (analytical and numerical) we make the following
conjecture:
\begin{quotation}
{\bf Conjecture.} For all rational $p/q$, we can find positive constant $c_{p /q}$, such that when
$\omega$ is in a small {\em real neighbourhood} of $p/q$, then
\begin{equation*}
\epscrit \sim c_{p /q} \Big\lvert \omega - \frac p q \Big\rvert ^{\frac 1 q}.
\end{equation*}
\end{quotation}

\indent
Comparing this conjecture with \cite{BerrettiGentile2,BerrettiMarmi} we see that the domain of
analyticity with respect to $\epsilon \in \mathbb{C}$ of the function defined by the series
\eqref{Lindstedt} is quite different from a circle as $\omega$ tends to a
resonance~\cite{berrettichierchia1,BCCF}. It is an open question to understand completely its
geometry: it could be an ellipse whose major semi--axis is in the direction of real $\epsilon$ with
a length proportional to the square root of the minor semi--axis, or probably a more complicated
curve with the nearest singularity to the origin placed on the imaginary axis. We don't think our
result and our conjecture are in disagreement with the one of \cite{BerrettiGentile2,BerrettiMarmi}
in fact we think there is a difference when dealing with real or complex frequencies.

\indent
{\it Acknowledgements.} 
We are grateful to S. Marmi and D. Sauzin 
for the important help they provided during this work and to B. Chirikov and
D. Shepelansky for some early discussions. Support from EC contract ERBCHRXCT94-0460 
is also acknowledged.

\section{Scaling law for the fundamental resonance.}
\label{sec:main}

The goal of this section is to prove our main result: Theorem~\ref{the:maintheorem}. We note that
from the definitions of $\epscrit$ and $\omega^{\pm}\left(\epsilon\right)$ we have
$\epsilon_{crit}\left(\omega^{\pm}\left(\epsilon\right)\right)=\epsilon$ then
Theorem~\ref{the:maintheorem} is equivalent to

\begin{theorem}
\label{the:secondvers}
  Let $\omega^{\pm}\left(\epsilon\right)$ be the functions of $\epsilon$ defined
in~\eqref{def:omegap} and ~\eqref{def:omegam}. Then there exist $\epsilon_0^{\prime}>0$ and
$c_0^{\prime}>0$ such that for all $0<\epsilon < \epsilon^{\prime}_0$ we have
\begin{equation*}
  \label{eq:boundfreq}
\Big \lvert \ln\lvert \omega^{\pm}\left(\epsilon\right)\rvert-\ln\epsilon \Big\rvert \leq
c_0^{\prime}.
\end{equation*}
\end{theorem}

\indent
Using Theorem $3.$ page $76$ of~\cite{TreschevZubelevich} we can prove that there exist $c>0$ and
$\epsilon_0>0$ such that for all $0<\epsilon < \epsilon_0$ there exist two invariant homotopically
non--trivial $\mathcal{C}^1$ curves with frequencies $\omega_{l}$ and $\omega_{b}$ and
\begin{equation*}
 \omega_{l} < 0 < \omega_{b} \quad \text{and} \quad \lvert \omega_{b} - \omega_{l}\rvert < c
\epsilon.
\end{equation*}
Recalling~\eqref{def:omegap} and \eqref{def:omegam} we get the lower bound
\begin{equation*}
  \label{eq:lowerbound}
  \ln \epsilon -\ln \lvert \omega^{\pm}\left(\epsilon\right)\rvert > \ln2 -\ln c
\end{equation*}
for $\epsilon$ small enough. Thus the proof of Theorem~\ref{the:secondvers} reduces to prove the
following bound
\begin{equation}
\label{eq:thebound}
  \ln \epsilon -\ln \lvert \omega^{\pm}\left(\epsilon\right)\rvert < c^{\prime\prime}
\end{equation}
for some $c^{\prime\prime}>0$ and $\epsilon$ small enough. This will be done in
Proposition~\ref{pro:finalestimate}. Now we present the two results we need to prove
Proposition~\ref{pro:finalestimate}.

\indent
Let us fix $\epsilon >0$ and recall the definitions of $y^{\pm}\left(\epsilon\right)$: $y^+\left(
\epsilon\right)$, respectively $y^-\left( \epsilon\right)$, are the intersection point of the
homotopically invariant $\mathcal{C}^1$ curve of frequency $\omega^+\left( \epsilon\right)$,
respectively $\omega^-\left( \epsilon\right)$, with the $y$--axis. Then from the exponentially
smallness of the splitting of the separatrices we deduce
\begin{proposition}
\label{pro:initialdata}
There exist $c>0$ and $\epsilon_0^{\prime\prime}>0$ such that for all $0<\epsilon <
\epsilon_0^{\prime\prime}$ we have
\begin{equation}
\label{eq:lastinvtori}
\lvert y^{\pm}\left( \epsilon\right) \rvert \geq c e^{-\frac{\pi^2}{2\sqrt{\epsilon}}}.
\end{equation}
\end{proposition}

We leave the proof of this Proposition to subsection~\ref{splitting} and we introduce the second
element we need to prove our result. We rewrite the Standard map as perturbation of the identity
map, namely
\begin{equation}
S_{\mu}\binom{x}{z}=\binom{x}{z}+\mu\binom{z}{\sin x}+\mu^2 \binom{\sin x }{0}
\label{standardmapidentity}
\end{equation}
we can pass from \eqref{standardmapidentity} to \eqref{standardmap} putting $\mu = \sqrt{\epsilon}$
and $\mu z=y$. And we define $\omega_S$ to be its rotation number, namely
\begin{equation*}
  \omega_S\left( x,z,\mu \right) = \lim_{n \rightarrow \infty} \frac{\pi_1\left(S^n_{\mu} \left( x,z
\right)\right) - x}{n}
\label{eq:rotnumsmid}
\end{equation*}
when this limit exists. The relation between $\omega_S$ and $\omega_T$ is 
\begin{equation}
\omega_T\left( x,y,\epsilon \right)=\sqrt{\epsilon} \omega_S\left(
x,\frac{y}{\sqrt{\epsilon}},\sqrt{\epsilon} \right).
\label{eq:frequencies}
\end{equation}

\indent
Following~\cite{BenettinGiorgilli} we construct (Proposition~\ref{pro:interpoling} and
Proposition~\ref{pro:expcloseflow}) an integrable Hamiltonian system whose time--$1$ flow differs
from the standard map in the formulation~\eqref{standardmapidentity} for an exponentially small
term: $\bigo{e^{-\frac{1}{\sqrt{\epsilon}}}}$ for $\epsilon \rightarrow 0^+$. This allows us to
prove (Proposition~\ref{prop:asymptoticexp}) the existence of an asymptotic development of the
standard map rotation number in terms of $\sqrt{\epsilon}$, whose first order term is the pendulum
frequency for rotation orbits. In fact the best interpolating Hamiltonian system is an integrable
perturbation of order $\bigo{\sqrt{\epsilon}}$ of the pendulum $H_0\left( x,z
\right)=\frac{z^2}{2}+\cos x$. This implies the following Lemma
\begin{lemma}
\label{lem:lemma1}
Let $h_0$ be a real number strictly greater than $1$. Let $\omega_T$ be the standard map rotation
number and let $\omega_{pend}$ be the frequency of a rotation orbit of energy $h_0$ for the pendulum
$H_p\left(x,z\right)=\frac{z^2}{2}+\cos x$. For $\epsilon >0$ we consider the invariant circle of
the standard map of initial conditions $\left( x_0,y_0\right)=\left(
x_0,\frac{y_0}{\sqrt{\epsilon}}\right)$, then for $\epsilon$ sufficiently small we have:
\begin{equation*}
\Big\lvert \omega_T\left( x_0,y_0,\epsilon \right) - \sqrt{\epsilon}\omega_{pend}\left(
x_0,\frac{y_0}{\sqrt{\epsilon}}\right) \Big\rvert = \bigo{\lvert \epsilon\rvert^{\frac 3 2}}.
\end{equation*}
\end{lemma}

We can now prove the bound~\eqref{eq:thebound}.

\begin{proposition}[Upper bound.]
\label{pro:finalestimate}
 Let $\omega^{\pm}\left(\epsilon\right)$ be the functions of $\epsilon$ defined
in~\eqref{def:omegap} and ~\eqref{def:omegam}. Then there exist $\bar{\epsilon}>0$ and
$c^{\prime\prime}>0$ such that for all $0<\epsilon < \bar{\epsilon}$ we have
\begin{equation*}
\ln \epsilon - \ln\lvert \omega^{\pm}\left(\epsilon\right)\rvert \leq c^{\prime\prime}.
\end{equation*}
\end{proposition}

\proof

Lemma~\ref{lem:lemma1} and its refinement (Corollary~\ref{pro:analyticprop}) say that the standard
map rotation number is ``well approximated'' by the pendulum frequency. Consider a rotation orbit of
the pendulum starting at $\left( 0,z_0\right)$. Its frequency is given by
\begin{equation*}
\label{eq:pendulum}
\omega_{pend}\left( h_0 \right) = \frac{\pi}{k} \frac{1}{\mathcal{K}\left( k \right)}
\end{equation*}
where $k^2=\left( 1+\frac{z_0^2}{4} \right)^{-1}$, $h_0 = 1+\frac{z_0^2}{2}$ and
\begin{equation*}
\mathcal{K}\left( k \right) = \int_0^{\frac{\pi}{2}}\frac{d\psi}{\sqrt{1-k^2 \sin ^2 \psi}}
\end{equation*}
is the complete Elliptic integral of first kind. For $k=1$ (that is $z_0=0$) the integral diverges
logarithmically and (see~\cite{abramowitzstegun}, formula $17.3.26$ p.$591$)
\begin{equation*}
  \lim_{z_0 \rightarrow 0}\left[\mathcal{K}-\frac{1}{2}\ln \frac{4+z_0^2}{z_0^2}\right]=2\ln2.
\end{equation*}

\indent
Therefore in the limit $z_0 \rightarrow 0$, $\omega_{pend}$ behaves as
\begin{equation}
\Big \lvert \omega_{pend} -\frac{\pi}{\ln\frac{2}{\lvert z_0 \rvert}}\Big \rvert =
\bigo{\frac{z_0^2}{\ln \lvert z_0 \rvert}}.
\label{eq:pendscal}
\end{equation}

\indent
Assume now that $\omega_T\left( 0, y_0,\epsilon \right)$ exists for $\epsilon$ small enough, where
$y_0=z_0\mu$. By ~\eqref{eq:frequencies}, Lemma~\ref{lem:lemma1} and \eqref{eq:pendscal} one has
\begin{equation}
\label{eq:quiteend}
\lvert \omega_T\left( 0,y_0,\epsilon \right) \rvert \geq \sqrt{\epsilon}
\frac{\pi}{\ln\frac{8\sqrt{\epsilon}}{\lvert y_0 \rvert}} + \bigo{\lvert \epsilon \rvert^{\frac 3
2}}.
\end{equation}
If $y_0$ is the intersection of the last invariant torus with the axis $x=0$, namely
$y_0=y^+\left(\epsilon\right)$ and $\omega_T\left( 0,y_0,\epsilon
\right)=\omega^+\left(\epsilon\right)$, then the bound~\eqref{eq:lastinvtori} gives 
\begin{equation*}
\lvert \omega^+\left(\epsilon\right) \rvert \geq \sqrt{\epsilon} \frac{\pi}{C+\frac 1 2 \ln
\epsilon+\frac{\pi^2}{2\sqrt{\epsilon}}}+\bigo{\lvert \epsilon\rvert^{\frac 3 2}}
\end{equation*}
for some positive constant $C$. Then for $\epsilon$ small enough we conclude that there exist
another positive constant $c$ such that
\begin{equation*}
\lvert  \omega^+\left(\epsilon\right) \rvert \geq c \epsilon.
\end{equation*}
Recalling that $\omega^-\left(\epsilon\right)=-\omega^+\left(\epsilon\right)$, the Proposition is
proved.
\endproof

\subsection{Splitting of the separatrices.}
\label{splitting}

The aim of this subsection is to introduce some results about the splitting of separatrices for the
Standard Map necessary to prove Proposition~\ref{pro:initialdata}. The splitting of the separatrices
has been studied by many authors, here we use the notations and the presentation given in
\cite{Gelfreich}.

\indent
For $\epsilon=0$ the circle $y=0$ is formed by fixed points of the Standard Map, for $\epsilon>0$
only two fixed points survive: $\left( 0,0 \right)$ and $\left( \pi,0 \right)$. The first one is
hyperbolic (for these values of $\epsilon$) and the other one elliptic. The linearised map at the
origin is
\begin{equation*}
  \begin{pmatrix}
  1+\epsilon & 1 \\ \epsilon & 1   
  \end{pmatrix}
\end{equation*}
and its eigenvalues are $\lambda$ and $\lambda^{-1}$ where
\begin{equation*}
  \lambda = 1 + \frac{\epsilon}{2}+\sqrt{\epsilon}\sqrt{1+\frac{\epsilon}{4}}.
\end{equation*}

\indent
The stable, $\mathcal{W}^s$, and unstable, $\mathcal{W}^u$, manifolds of this fixed point are
analytic curves passing through $\left( 0,0 \right)$. We introduce the parameter $\delta=\ln
\lambda$ and we note that $\epsilon \sim \delta^2$ for small $\epsilon$. We parametrise the part of
the unstable, $\mathcal{W}^u_1$ separatrix which grows upwards from the origin, with $\left(
x^-\left( t \right),y^-\left( t \right) \right)$, imposing the boundary conditions\footnote{The
$y\left( t \right)$ can be found using $y\left( t+\delta \right)=x\left( t+\delta \right) - x\left(
t \right)$.}
\begin{equation*}
  \lim_{t \rightarrow -\infty}x^{-}\left( t \right)=0, \quad x^{-}\left( 0 \right)=\pi.
\end{equation*}
We assume that $t=0$ is the first intersection of $\mathcal{W}^u_1$ with the vertical line $x=\pi$.
The parametrisation of $\mathcal{W}^s_1$ is given by
\begin{equation*}
  \left( x^+\left( t \right), y^+\left( t \right) \right)=\left( 2 \pi- x^-\left( -t \right),
y^-\left(-t \right)+\epsilon \sin x^-\left( -t \right) \right).
\end{equation*}
One can show that
\begin{equation*}
  \lim_{t \rightarrow +\infty}x^{+}\left( t \right)=0, \quad x^{+}\left( 0 \right)=\pi  ,
\end{equation*}
i.e. $t=0$ corresponds to a homoclinic point. Lazutkin~\cite{Lazutkin} proposed to study the {\em
homoclinic invariant} defined by
\begin{equation*}
  \Omega = \det 
  \begin{pmatrix}
    \dot x^-\left( 0 \right) & \dot x^+\left( 0 \right) \\
         \dot y^-\left( 0 \right) & \dot y^+\left( 0 \right)
  \end{pmatrix}.
\end{equation*}
The homoclinic invariant has the same value for all points of one homoclinic trajectory and it is
invariant with respect to symplectic coordinate changes. Gelfreich in \cite{Gelfreich}, Proposition
$2.1$, gives an asymptotic expansion in power of $\delta$ of the parametrisation $x^{\pm}\left( t
\right)$. Its first terms are
\begin{gather}
  x_0\left( t \right)+x_1\left( t \right)+x_2\left( t \right)+\cdots = \notag \\
  =4 \arctan e^t+\delta^2 \frac{1}{4} \frac{\sinh t}{\left( \cosh t \right)^2}+\delta^4 \left[-
\frac{41}{1728} \frac{\sinh t}{\left( \cosh t \right)^2}+ \frac{91}{864} \frac{\sinh t}{\left( \cosh
t \right)^4}\right]+\cdots .
\label{formsol}
\end{gather}
Of course the zero order term is nothing but the parametrisation of the separatrix of the pendulum.
\begin{center}
  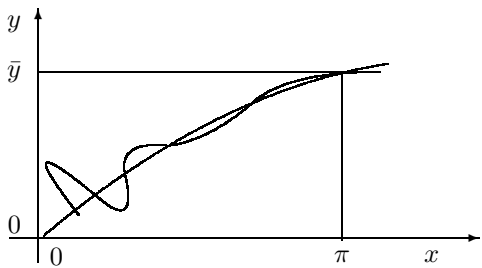
\begin{figure}[ht]
   \makebox{
   \unitlength=1.000000pt
    \begin{picture}(182.00,122.00)(0.00,0.00)
      \put(130,72){\line(-1,0){115}}
      \put(6,72){\makebox(0.00,0.00){$\bar{y}$}}
      \put(130.00,2.00){\makebox(0.00,0.00){$\pi$}}
      \put(6.00,90.00){\makebox(0.00,0.00){$y$}}
      \put(164.00,2.00){\makebox(0.00,0.00){$x$}}
      \put(6.00,14.00){\makebox(0.00,0.00){$0$}}
      \put(22.00,2.00){\makebox(0.00,0.00){$0$}}
      \put(130.00,72.00){\line(0,-1){64.00}}
      \qbezier(98.00,62.00)(112.00,72.00)(144.00,72.00)
      \qbezier(48.00,32.00)(44.00,46.00)(64.00,44.00)
      \qbezier(38.00,24.00)(2.00,54.00)(30.00,18.00)
      \qbezier(38.00,24.00)(52.00,12.00)(48.00,32.00)
      \qbezier(98.00,62.00)(80.00,46.00)(64.00,44.00)
      \qbezier(17.00,10.00)(82.00,65.00)(147.00,75.00)
      \put(15.00,0.00){\vector(0,1){96.00}}
      \put(4.00,9.00){\vector(1,0){178.00}}
    \end{picture}
    }  
    \caption{The stable and unstable manifolds for the Standard Map with some of their infinitely
many intersection points.}
    \label{splitt1}
  \end{figure}
\end{center}

\indent
The main result of \cite{Gelfreich} is $\Omega \geq \frac{c}{\delta^2}e^{-\frac{\pi^2}{\delta}}$ for
some positive constant $c$ and $\delta$ small enough. Using this result we obtain an estimation of
$y^{\pm}\left( \epsilon\right)$ in Proposition~\ref{pro:initialdata}.

\indent
{\bf Proposition~\ref{pro:initialdata}.} {\em There exist $c>0$ and $\epsilon_0^{\prime\prime}>0$
such that for all $\epsilon < \epsilon_0^{\prime\prime}$ we have}
\begin{equation*}
\lvert y^{\pm}\left( \epsilon\right) \rvert \geq c e^{-\frac{\pi^2}{2\sqrt{\epsilon}}}.
\end{equation*}

\begin{remark}
Compare with Figure~\ref{logyvsefig} where we plot $\ln \rvert y^+\left( \epsilon_{crit}
\right)-y^-\left( \epsilon_{crit} \right)\rvert$ against $\epsilon_{crit}$, with data obtained
numerically using the FMA.
\end{remark}

\proof
For a fixed value of $\epsilon >0$, the stable and unstable manifolds intersect firstly at $\left(
\pi, \bar{y}\left(\epsilon\right) \right)$ and then infinitely many times. We call {\em lobe} the
arc of manifold between two successive intersections and {\em lobe area} the area bounded by the
lobe. Each lobe will be distinguished by  an integer: the number of iterates (of $T_{\epsilon}$)
needed to put the point $\left( \pi, \bar{y} \right)$ onto the right point base of the lobe.

\indent
The hyperbolic fixed point $\left( 0,0 \right)$ stretches the unstable manifold along the dilating
direction (corresponding to the eigenvalue $\lambda$), so there exists an integer $\hat{k}$ such
that the $\hat{k}$-th lobe intersect the $y$--axis at some point $y_{int}\left(\epsilon\right)$.
Because the invariant circles of the Standard Map are at least Lipschitz graph, we cannot have
homotopically non--trivial invariant circles, passing by $\left(0,\bar{y}^{\prime} \right)$ with
$\bar{y}^{\prime} < y_{int}\left(\epsilon\right)$, thus $y^{+}\left( \epsilon\right) \geq
y_{int}\left(\epsilon\right)$. The lobe area is conserved under iteration of the Standard Map and by
\cite{Gelfreich}, Corollary~$1.3$, this area is\footnote{In the follow we will use often the
standard notation (see~\cite{Godement} n.$4$ chap.~II) $f\left(t\right)\sim g\left(t\right)$ for
$t\rightarrow 0$, this means that $f$ and $g$ are equivalent when $t$ is close enough to $0$. We can
restate it by saying: $f\left( t \right)=g\left(t\right)+{\it o}\!\left(g\left(t\right)\right)$ when
$t\rightarrow 0$.} for $\delta$ small enough
\begin{equation*}
  \label{lobearea}
  A_{lobe}\sim\frac{2\omega_0}{\pi} e^{-\frac{\pi^2}{\delta}}
\end{equation*}
where $\omega_0=1118.827706\cdots$ and $\delta=\ln \lambda \sim \sqrt{\epsilon}$ for $\epsilon$
small.

\indent
In Figure~\ref{kiter} we represent the geometry of the generic $k$--th lobe: the points $A_k=\left(
x_{-k},y_{-k}\right)$ and  $B_k=\left( x_{-(k-1)},y_{-(k-1)}\right)$ are the iterates of the
standard map
\begin{equation*}
  \begin{cases}
    x_{-(k-1)}=x_{-k}+y_{-(k-1)} \\
    y_{-(k-1)}=y_{-k}+\delta^2\sin x_{-k}
  \end{cases}
\end{equation*}
the segment of line $\overline{A_kC_k}$ is tangent to the unstable manifold at $A_k$, while
$\overline{A_kB_k}$ is tangent to the stable manifold at $A_k$.
\begin{center}
  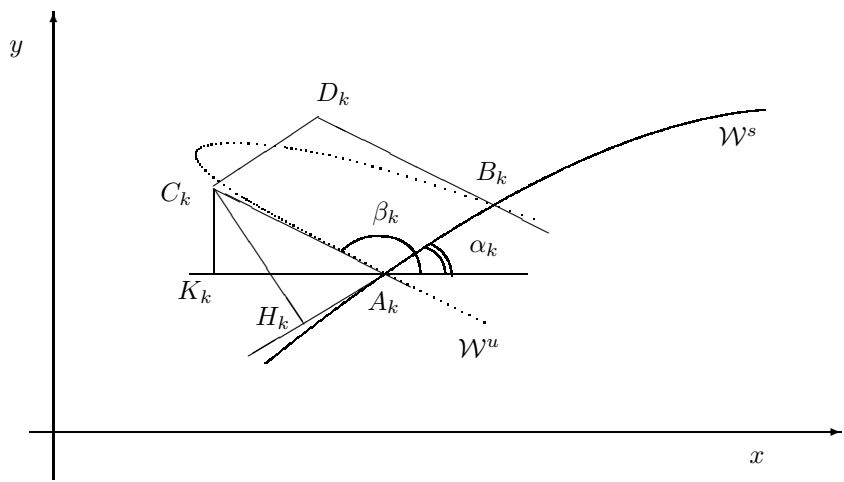
\begin{figure}[ht]
   \makebox{
\unitlength=1.000000pt
\begin{picture}(312.00,177.00)(0.00,0.00)
\put(139.50,77.75){\line(-5,-3){51.75}} 
\put(97,61.00){\makebox(0.00,0.00){$H_k$}}
\put(67.50,71.00){\makebox(0.00,0.00){$K_k$}}
\put(74.75,110.00){\line(2,-3){34.00}} 
\put(74.75,109.75){\line(0,-1){32.25}} 
\put(139.00,67.00){\makebox(0.00,0.00){$A_k$}}
\put(180.00,116.00){\makebox(0.00,0.00){$B_k$}}
\put(120.00,146.00){\makebox(0.00,0.00){$D_k$}}
\put(60.50,108.00){\makebox(0.00,0.00){$C_k$}}
\qbezier(138.75,92.00)(130.00,93.25)(123.00,86.75)
\put(140.00,101.00){\makebox(0.00,0.00){$\beta_k$}}
\qbezier(152.50,77.75)(152.25,90.50)(138.50,92.25)
\put(177.00,87.50){\makebox(0.00,0.00){$\alpha_k$}}
\qbezier(164.50,77.00)(164.00,85.25)(156.50,89.25)
\qbezier(162.00,77.75)(161.50,85.50)(154.00,88.00)
\put(193.25,77.75){\line(-1,0){128.00}} 
\put(0.00,163.00){\makebox(0.00,0.00){$y$}}
\put(280.00,8.50){\makebox(0.00,0.00){$x$}}
\put(114.25,137.50){\line(-3,-2){39.75}} 
\put(201.25,93.50){\line(-2,1){87.00}}  
\put(149.00,73.00){\line(-2,1){74.50}} 
\qbezier[30](102.00,125.50)(126.25,123.25)(195.50,98.50)
\qbezier[30](85.75,106.50)(42.75,132.75)(101.25,126.00)
\qbezier[50](176.67,59.33)(105.00,95.00)(85.00,107.00)
\put(175,50){\makebox(0,0){$\mathcal{W}^u$}}
\qbezier(94.00,44.00)(206.00,134.00)(283.00,140.00)
\put(273,130){\makebox(0,0){$\mathcal{W}^s$}}
\put(14.00,0.00){\vector(0,1){177.00}}
\put(5.00,18.00){\vector(1,0){307.00}}
\end{picture}
    }  
    \caption{Two intersection points of the stable, $\mathcal{W}^s$, and unstable, $\mathcal{W}^u$
(dotted curve), manifolds, producing the $k$--th lobe.}
    \label{kiter}
  \end{figure}
\end{center}
The angle $\alpha_k$ is defined by
\begin{equation*}
  \tan \alpha_k=\frac{y_{-(k-1)}-y_{-k}}{x_{-(k-1)}-x_{-k}}=\frac{\delta^2\sin x_{-k}}{\Delta
x_{-k}}.
\end{equation*}

\indent
Let us consider the lobe which intersects for the first time the $y$--axis and let us call $y_{int}$
the intersection point. We don't know exactly the analytical form of the arc of unstable manifold
joining $A_{\Hat{k}}=\left( x_{-\Hat{k}},y_{-\Hat{k}}\right)$ to $\left(0,y_{int}\right)$, so we can
not calculate exactly $y_{int}$. Then we approximate this arc with its tangent at $A_{\Hat{k}}$:
\begin{equation}
  \label{eq:tanline}
  y=y_{-\Hat{k}}+\left(\tan \beta_{\Hat{k}}\right)\left( x-x_{-\Hat{k}}\right).
\end{equation}
This straight line intersect the $y$--axis in some point $\tilde{y}$ and $\lvert
\tilde{y}-y_{int}\rvert = \bigo{x_{-\hat{k}}^2}$.

\indent
We approximate the lobe with a parallelogram of sides $A_kB_k$ and $A_kC_k$ where:
\begin{enumerate}
\item the straight line through $A_k$ and $C_k$ is tangent to the lobe at $A_k$;
\item the area, $\overline{A_kB_k}\cdot \overline{A_kC_k}\sin\left( \beta_k-\alpha_k\right)$, of the
parallelogram is equal to $A_{lobe}$.
\end{enumerate}
The condition of intersection with the $y$--axis is now
$\overline{A_{\Hat{k}}K_{\Hat{k}}}=x_{-\Hat{k}}$: this defines implicitly $\Hat{k}$.

\indent
Let $H_k$ denote the orthogonal projection of $C_k$ on the straight line through $A_k$ and $B_k$.
Then $A_{lobe}=\overline{A_kB_k} \cdot \overline{C_kH_k}=\frac{\Delta x_{-k}}{\cos
\alpha_k}\overline{C_kA_k} \sin\left( \beta_k - \alpha_k\right)$. Since
$\overline{A_kK_k}=-\overline{C_kA_k}\cos \beta_k$, the intersection condition becomes
\begin{equation}
  \overline{A_{\Hat{k}}K_{\Hat{k}}}=\frac{-\cos \alpha_{\Hat{k}}\cos \beta_{\Hat{k}}}{\Delta
x_{-\Hat{k}}\sin\left( \beta_{\Hat{k}} -\alpha_{\Hat{k}}\right)}A_{lobe}=x_{-\Hat{k}}.
\label{eq:interscond}
\end{equation}

\indent
From~\eqref{formsol}, using $t=-\Hat{k}\delta$ one gets
\begin{equation*}
x_{\hat{k}}\sim 4e^{-\hat{k}\delta}, \quad \Delta x_{-\hat{k}}\sim 4\delta e^{-\hat{k}\delta},
\end{equation*}
thus
\begin{equation*}
\tan \alpha_{\hat{k}} \sim \delta\left( 1+e^{-2\hat{k}\delta}\right), \quad \sin \alpha_{\hat{k}}
\sim \delta\left( 1+e^{-2\hat{k}\delta}\right), \quad \cos \alpha_{\hat{k}} \sim 1.
  \label{step2}
\end{equation*}
To estimate $\beta_{\hat{k}}$ we use again the fact that the intersection will happen very close to
the origin: the tangent to the unstable manifold at $A_{\hat{k}}$ is very close to the expanding
direction at the origin, so we get
\begin{equation*}
  \tan \beta_{\hat{k}}\sim -\delta\left(1+\frac \delta 2\right), \quad \cos \beta_{\hat{k}}\sim
1-\delta^2.
\end{equation*}

Using these approximations we can rewrite~\eqref{eq:interscond} as
\begin{equation*}
\left(1-\delta^2\right)\frac{2\omega_0}{\pi}e^{-\frac{\pi^2}{\delta}}=4e^{-\hat{k}\delta}\frac{4\delta
e^{-\hat{k}\delta}}{1+e^{-2\hat{k}\delta}}2\delta +{\it o}\left( \delta^2e^{-2\hat{k}\delta}\right),
\end{equation*}

from which we obtain
\begin{equation*}
  \delta^2e^{-2\hat{k}\delta}=C_1e^{-\frac{\pi^2}{2\delta}}+{\it o}\left(
e^{-\frac{\pi^2}{\delta}}\right)
\label{hatk}
\end{equation*}
for some positive constant $C_1$. This allows to compute $\hat{k}$ explicitly. To determine
$\tilde{y}$, we recall \eqref{eq:tanline} and using the previous value of $\hat{k}$ we get
\begin{equation*}
 \tilde{y} = y_{-\hat{k}}-x_{-\hat{k}}\tan \beta_{\hat{k}} \sim 4\delta
e^{-\hat{k}\delta}+4e^{-\hat{k}\delta}\delta\left( 1+\frac \delta 2\right) \sim
C_1^{\prime}e^{-\frac{\pi^2}{2\delta}},
\label{yintdef}
\end{equation*}
for some new positive constant $C_1^{\prime}$. Finally, since $\lvert \tilde{y}-y_{int}\rvert
=\bigo{x^2_{-\hat{k}}}=\bigo{e^{-\frac{\pi^2}{\sqrt{\epsilon}}}}$, we conclude
\begin{equation*}
-y^{-}\left( \epsilon\right)=y^{+}\left( \epsilon\right) \geq  ce^{-\frac{\pi^2}{2\sqrt{\epsilon}}},
\end{equation*}
for some positive constant $c$.
\endproof

\subsection{Interpolating vector fields.}
\label{interpoling}

Once we have determined a lower bound for $y^+\left(\epsilon\right)$ in terms of the intersection of
the lobe with the $y$--axis, we need to compute a lower bound for the rotation number of the
corresponding invariant circle. To this purpose following~\cite{BenettinGiorgilli}, we consider the
problem of the construction of a vector field interpolating a given diffeomorphism. With this
subsection and the following we prove the existence of an interpolating Hamiltonian system
exponentially close to the map~\eqref{standardmapidentity} (Proposition~\ref{pro:interpoling} and
Proposition~\ref{pro:expcloseflow}). These results will be used in section~\ref{ssec:asymptdev} to
prove the existence of an asymptotic development for the standard map rotation number
(Proposition~\ref{prop:asymptoticexp}) from which Lemma~\ref{lem:lemma1} immediately follows.

\indent
Let $D$ be a bounded subset of $\R^m$ for $m \in \N^*$, $I$ a real interval containing the origin
and let us consider the one--parameter family of diffeomorphisms $\Psi$
\begin{align*}
  \Psi :& D \times I \rightarrow \R^m \\
       & \left(x , \mu \right) \mapsto  \Psi \left(x ,\mu\right)
\end{align*}
such that for all $x \in D$ and $\mu \in I$ it has the form
\begin{equation}
  \label{eq:diffeo}
  \Psi \left( x,\mu \right) = x + \sum_{n \geq 1 } \Psi_n\left( x \right) \mu^n.
\end{equation}

\indent
The problem consists in finding a one--parameter family of $F:D \times I \rightarrow \R^m$ vector
fields such that the time--$1$ flow of the differential equation
\begin{equation*}
\frac{d x}{d t} = F\left( x,\mu \right)
\label{eq:edo}
\end{equation*}
coincides with the diffeomorphism $\Psi$.

\indent
We assume that $\Psi_n\left( x \right)$, defined in \eqref{eq:diffeo}, are real analytic functions
of $x$ in some complex neighbourhood $\D_{\rho}$ of $D$, defined as follows. Given $\rho = \left(
\rho_1, \ldots ,\rho_m \right) \in \R_+^m$ then
\begin{equation*}
  \label{eq:cmplx}
  \D_{\rho}=\bigcup_{x \in D} \{ z \in \C^m: \lvert z_i - x_i \rvert < \rho_i, 1\leq i \leq m \}.
\end{equation*}

\indent
We introduce standard norms on complex analytic functions. Let $f$ be a scalar analytic function on
some domain $D \subset \R^m$, let $\rho \in \R_+^m$ and consider its complex extension $\D_{\rho}$,
then
\begin{equation}
  \normm{f}{\rho}=\sup_{z \in \D_{\rho}} \lvert f\left( z \right) \rvert
\label{eq:defnormsc}
\end{equation}
defines a norm. If $F$ is vector valued analytic function $F=\left( f_1,\dots,f_m \right)$ on
$\D_{\rho}$, for some $\rho=\left( \rho_1,\dots ,\rho_m \right) \in \R^m_+$, then we define a norm
(noted as in the scalar case)
\begin{equation*}
  \normm{F}{\rho}=\max_{1 \leq i \leq m} \normm{f_i}{\rho}.
\label{eq:defnormve}
\end{equation*}

\indent
If $\Psi$ is a formal $\mu$--series we can look for a formal interpolating vector field $F\left(
x,\mu \right) = \sum_{n \geq 1}F_n\left( x \right) \mu^n$ whose time--$1$ flow coincides with the
diffeomorphism, as we show with the following Proposition.

\begin{proposition}
  \label{prop:formal}
Let $\Psi\left( x,\mu \right)$ be a diffeomorphism of the form \eqref{eq:diffeo} and assume $\Psi_n
\left( x \right)$ be analytic in $\D_{\rho}$ for all $n\geq 1 $ and for some $\rho \in R_+^m$. Then
there exists a formal vector field $F_{\mu}$ of the form
\begin{equation}
  \label{eq:formvf}
F_{\mu}\left( x \right)=\sum_{n \geq 1 }F_n\left( x \right) \mu^n
\end{equation}
such that
\begin{equation}
  \label{eq:timeone}
e^{L_{F_{\mu}}}\left( x \right) = \Psi \left( x,\mu \right)
\end{equation}
for all $x \in \D_{\rho}$.
\end{proposition}

\proof
The proof is a very natural application of the Lie series. We give it for the sake of completeness.

\indent
By definition of exponential of the Lie derivative
\begin{equation*}
  e^{L_{F_{\mu}}}\left( x \right) = x + \sum_{n \geq 1 }\frac{1}{n!}L^n_{F_{\mu}}\left( x \right).
\end{equation*}
Looking for $F_{\mu}$ in the form \eqref{eq:formvf} and using the linearity of the Lie operator we
have
\begin{equation*}
  L_{F_{\mu}} = \sum_{n \geq 1}\mu^n L_{F_n}.
\end{equation*}

\indent
For all integer $n \geq 1$ and $2 \leq m \leq n$ we define
\begin{equation*}
  \label{eq:lmn}
\L_{1,n}\left( x \right)=L_{F_n}\left( x \right)=F_n\left( x \right) \quad \L_{m,n}\left( x
\right)=\sum_{j=1}^{n-m+1}L_{F_j}\L_{m-1,n-j}\left( x \right)
\end{equation*}
and it is easy to prove that
\begin{equation*}
  L^m_{F_{\mu}}=\sum_{n \geq m} \mu^n \L_{m,n}.
\end{equation*}
Then by \eqref{eq:timeone} and  \eqref{eq:diffeo} we conclude that for all $k\geq 1$
\begin{equation*}
  \sum_{n=1}^k \frac{1}{n!}\L_{n,k}\left( x \right)=\Psi_k\left( x \right),
\end{equation*}
namely
\begin{equation}
  L_{F_k}\left( x \right)=\Psi_k\left( x \right)-\sum_{n=2}^k \frac{1}{n!}\L_{n,k}\left( x \right).
\label{eq:formsol2}
\end{equation}
Let $\Psi_k = \left( \Psi^{(1)}_k,\dots, \Psi^{(m)}_k \right)$ and let
$G=\sum_{n=2}^{k}\frac{1}{n!}\L_{n,k}\left( x \right) = \left( G_1,\dots, G_m\right)$. Note that $G$
depends on $\Psi_j$ and $F_{j-1}$ for $j \leq k$. Using \eqref{eq:formsol2} we have for all $1 \leq
l \leq m$
\begin{equation*}
F^{\left( l \right)}_k\left( x \right)  = \Psi^{(l)}_k+G_l,
\label{eq:formsol}
\end{equation*}
from which we obtain $F_k\left( x \right)=\left( F^{\left( 1 \right)}_k\left( x \right),\ldots,
F^{\left( m \right)}_k\left( x \right) \right)$.
\endproof

\indent
If $\Psi$ is analytic w.r.t $\mu$ then we have the following result(~\cite{BenettinGiorgilli}
Proposition $1$ p.$1122$)

\begin{theorem}
  \label{the:thetheorem}
Let $\Psi \left(x,\mu\right)$ be a diffeomorphism of the form \eqref{eq:diffeo} with $\Psi_n\left( x
\right)$ real analytic in some complex domain $\D_{\rho}$ for $\rho \in \R_+^m$, and let $F_{\mu}$
be the formal interpolating vector field obtained in Proposition~\ref{prop:formal}. Assume that for
all $k$
\begin{equation*}
  \normm{\Psi_k}{\rho}\leq \Gamma \gamma^{k-1},
\label{eq:psigev}
\end{equation*}
for some positive constants $\Gamma$ and $\gamma$ then
\begin{equation*}
\normm{F_1}{\rho} \leq \Gamma, \quad \forall k\geq 2 \quad \normm{F_k}{\frac{\rho}{2}} <\frac 1 2
k^{k-1}\beta^{k-1} \Gamma,
\end{equation*}
with $\beta=4 \max \{ \gamma, \Gamma \}$.

\indent
Moreover if $\Psi \left( x,\mu \right)$ is symplectic then for all $n$ $F_n\left( x \right)$ are
locally Hamiltonian.
\end{theorem}

\indent
In the form~\eqref{standardmapidentity} the Standard map verifies the hypothesis of
Theorem~\ref{the:thetheorem}, but to get optimal estimates we state an Hamiltonian version of
Theorem~\ref{the:thetheorem} adapted to the Standard map.

\begin{proposition}
\label{pro:interpoling}
Let $\bar{z} >0$ and consider the interval $I=\left( -\bar{z},\bar{z} \right)$. Fixed $R_0 >0$ and
$S_0 > 0$ we define the complex cylinder
\begin{equation*}
\D_{(R_0,S_0)}=\{ x \in \C /2\pi \Z :\lvert \Im x \rvert < S_0 \} \times \bigcup_{z \in
I}\{z^{\prime} \in \C : \lvert z - z^{\prime} \rvert < R_0 \}.
\end{equation*}
Then we can find a formal Hamiltonian $H\left( x,z,\mu \right) = \sum_{n=1}^{\infty} H_{n-1}\left(
x,z\right) \mu^n$ such that
\begin{equation*}
\forall \left( x,z \right) \in \D_{(R_0,S_0)} \quad e^{L_{H}}\left( x,z \right) = S_{\mu} \left( x,z
\right).
\end{equation*}
Moreover for all $0 < d <1$ there exist $A=2S_0 \normm{H_0}{(R_0,S_0)}$ and $B=\frac{20}{3}
\frac{\normm{H_0}{(R_0,S_0)}}{R_0 d^2}$ such that
\begin{equation*}
\forall k \quad \normm{H_k}{\left(1-d\right)\left( R_0,S_0 \right)} \leq A B^k \left( k+1 \right) !,
\end{equation*}
namely $H$ is $1$--Gevrey w.r.t. $\mu$.
\end{proposition}

With norm defined in~\ref{eq:defnormsc} it results that: $\normm{H_0}{\left( R_0,S_0 \right)} =
\frac{R_0^2}{2}+e^{S_0}$.

\indent
The proof of Proposition~\ref{pro:interpoling} is an adaptation of the proof of
Theorem~\ref{the:thetheorem} and we omit it.

\indent
We write down the first few terms of $H$
\begin{equation}
  H\left( x,z,\mu \right)= \frac{z^2}{2}+\cos x + \mu \frac{z}{2} \sin x + \bigo{\mu^2}
\label{eq:firstterms}
\end{equation}
and as we stated this is a perturbation of order $\bigo{\sqrt{\epsilon}}$ of the pendulum.

\subsection{Exponentially small estimates.}
\label{expsmallest}

Proposition~\ref{pro:interpoling} gives us an Hamiltonian function represented as a (maybe
divergent) series. It is then interesting to define a truncation of this series, namely take $N \in
\N^*$ define $H^{\left( N \right)}\left( x,z,\mu \right)$ by
\begin{equation*}
  \label{eq:tronc}
  H^{\left( N \right)}\left( x,z,\mu \right) = \sum_{n=1}^N H_{n-1}\left( x,z \right) \mu^n
\end{equation*}
and to ask for the relations between $S_{\mu}\left( x,z \right)$ and $\Psi^{\left( N \right)}\left(
x,z,\mu \right)$, where the latter is defined by
\begin{equation*}
  \Psi^{\left( N \right)}\left( x,z,\mu \right) = e^{L_{H^{\left( N \right)}}}\left( x,z \right)
\end{equation*}

\indent
We can prove that we can chose $N$ such that $S_{\mu}\left( x,z \right)$ and $\Psi^{\left( N
\right)}\left( x,z,\mu \right)$ are exponentially close. More exactly:

\begin{proposition}
\label{pro:expcloseflow}
Let $H^{\left( N \right)}\left( x,z,\mu \right)$ be the truncation of the $1$--Gevrey interpolating
Hamiltonian system given by Proposition~\ref{pro:interpoling} , and let $\Psi^{\left( N
\right)}\left( x,z,\mu \right)$ be its time--$1$ flow. Then we can find a positive integer $N^*$
such that for all $0 < d <\frac 2 3$ and for all $\lvert \mu \rvert \leq \frac{e R_0 S_0 d^2}{32 A}$
\begin{equation*}
\label{eq:expclose}
\normm{S_{\mu}\left( x,z \right)-\Psi^{\left( N^* \right)}\left( x,z,\mu \right)}{\left( 1 -
\frac{3}{2} d \right) \left( R_0,S_0 \right)} \leq \lvert \mu \rvert D e^{-\frac{D^{\prime}}{\lvert
\mu \rvert}},
\end{equation*}
with $D=M+\frac{16 A m }{eR_0 S_0 d^2}$, $D^{\prime}=\left( 2eB \right)^{-1}$, $M=\max\{ R_0+\lvert
\mu \rvert e^{S_0},e^{S_0} \}$ and $m=\max\{ R_0,e^{S_0} \}$.
\end{proposition}

\indent
This Proposition is an adaptation of the corollary $1$ p.$1124$ of \cite{BenettinGiorgilli} and we
don't prove it.

\indent
We will note in the follow $H^*\left( x,z,\mu \right)=H^{\left(N^*\right)}\left( x,z,\mu \right)$:
the ``best'' truncation of the interpolating Hamiltonian.

\indent
Let us introduce the frequency of the rotation orbit of $H^*$.
\begin{equation*}
2 \pi \omega_* \left( x,z,\mu \right)=\lim_{n \rightarrow \infty}\frac{\pi_1\left(
e^{nL_{H^*}}\left(x,z\right) \right)-x}{n}.
\label{eq:rotnumstar}
\end{equation*}
Then from Proposition~\ref{pro:expcloseflow} we deduce the following corollary.

\begin{corollary}
\label{pro:analyticprop}
Let $\omega_S$ be the rotation number of the map \eqref{standardmapidentity}. Let $H^*$ be best
truncation of the interpolating Hamiltonian system given by Proposition~\ref{pro:expcloseflow} and
let $2 \pi \omega_*$ be its frequency. Then we have
\begin{equation}
\label{eq:expsmallfreq}
\lvert  \omega_S\left( x,z,\mu \right) -  \omega_*\left( x,z,\mu \right) \rvert  \leq  D\lvert \mu
\rvert e^{-\frac{D^{\prime}}{\lvert \mu \rvert}}.
\end{equation}
\end{corollary}

\proof
We can prove that for all positive $k$ and for initial data $\left( x,z \right)$ on some domain such
that we can iterate $e^{L_{H^*}}$ and $S_{\mu}$ at least $k$ times, we have
\begin{equation*}
S_{\mu}^k\left( x,z \right) - e^{k L_{H^*}}\left( x,z \right) = \sum_{l=0}^{k-1} e^{(k-l-1) L_{H^*}}
\left( S_{\mu} - e^{L_{H^*}} \right) S^l_{\mu}\left( x,z \right) .
\end{equation*}

\indent
For real $x$, $z$ and $\mu$, the standard map is real valued so $\xi_l=S^l_{\mu}\left( x,z \right)$
is in $2\pi \T \times \R$ for all $l\geq 0$.

\indent
Let us define $\xi_l^{\prime}=\left( S_{\mu} - e^{L_{H^*}} \right)\left( \xi_l \right)$, then
Proposition~\ref{pro:expcloseflow} implies $\lvert \xi_l^{\prime} \rvert \leq D \lvert \mu \rvert
e^{-\frac{D^{\prime}}{\lvert \mu \rvert}}$ for $(x,z) \in \D_{(1-\frac 3 2 d)(R_0,S_0)}$. Such a $D$
depends on the domain troughs $d$, for some fixed $S_0$ and $R_0$, but using the previous remark on
the reality of $\xi_l$, we can fix once for all a domain, and so a constant $D$, in such a way that
$\xi_l$ belongs for all $l$ to  this domain. For all $n$: $\lvert e^{n L_{H^*}} \left(
\xi_l^{\prime} \right) \rvert \leq D \lvert \mu \rvert e^{-\frac{D^{\prime}}{\lvert \mu \rvert}}$,
because $e^{nL_{H^*}}\left( \xi_l^{\prime}\right)=\xi_l^{\prime}+\bigo{\xi_l^{\prime}}$.

\indent
We conclude that 
\begin{equation*}
\Big \lvert \left[ \pi_1\left(S^k_{\mu} \left( x,z \right)\right)  - x \right] - \left[
\pi_1\left(e^{k L_{H^*}} \left( x,z \right)\right) - x \right] \Big \rvert \leq  k \lvert \mu \rvert
D e^{-\frac{D^{\prime}}{\lvert \mu \rvert}},
\end{equation*}
from which \eqref{eq:expsmallfreq} follows.
\endproof

\subsection{Asymptotic development.}
\label{ssec:asymptdev}

We can now prove the existence of an asymptotic development for the standard map rotation number in
a neighbourhood of $\epsilon=0$.

\indent
The Hamiltonian $H^*$ defines an integrable (it has only one degree of freedom) Hamiltonian system
whose period is given by
\begin{equation}
\label{eq:integrale}
T=\frac{2 \pi}{\omega_*}= \int_0^{2\pi} \frac{dx}{\frac{\partial H^*}{\partial z}\Big \rvert_{h_0}},
\end{equation}
where $h_0$ is an energy level corresponding to rotation orbits.

\indent
Let us fix $h_0>1$. We are interested in finding curves on the energy level
$H^*\left(x,z,\mu\right)~=~h_0$. Take $z_0>0$ and consider the point $P_0=\left( x_0,z_0,0\right)$
such that $P_0 \in {H^*}^{-1} \{ h_0 \}$, from \eqref{eq:firstterms} it follows that:
$\frac{\partial H^*}{\partial z}\left( P_0 \right) = z_0 >0$. We can then apply the implicit
function Theorem which assures that it exists a function $\hat{z}=\hat{z}\left( x,\mu \right)$ such
that locally it satisfies
\begin{equation}
  H^*\left( x,\hat{z},\mu \right) = h_0
\label{eq:implicitfthe}
\end{equation}
moreover this function is smooth. We compute the first terms and we get: $\hat{z}\left( x, \mu
\right)= \sqrt{2 \left( h_0 -\cos x \right)}-\frac{\mu}{2}\sin x +\bigo{\lvert \mu \rvert ^2}$ and
\begin{align}
  \frac{\partial H^*}{\partial z}\left( x, \hat{z},\mu \right) &- \sqrt{2 \left( h_0 -\cos x
\right)} \notag \\
  &= \hat{z}+ \frac{\mu}{2}\sin x +\sum_{n=2}^{N^*} \frac{\partial H_n}{\partial z}\left( x,
\hat{z}\right)  \mu^n - \sqrt{2 \left( h_0 -\cos x \right)} =\bigo{\lvert \mu \rvert ^2}.
\label{eq:develop}
\end{align}
Recalling ~\ref{eq:frequencies} we prove the following Proposition

\begin{proposition}
  \label{prop:asymptoticexp}
Let $\omega_T\left( x,y,\epsilon\right)$ be the rotation number of the standard map in the
formulation~\eqref{standardmap}, then it exists a formal $\sqrt{\epsilon}$--power series
$\sum_{n=0}^{\infty}B_n \left(\sqrt{\epsilon}\right)^n$ asymptotic to $\omega_T\left(
x,y,\epsilon\right)$ for $\epsilon\rightarrow 0$. Namely for all positive integer $N$ we have
\begin{equation}
  \label{eq:asympexp}
 \lim_{\epsilon\rightarrow 0}\left(\sqrt{\epsilon}\right)^{-\left(N+1\right)}\Big\lvert \,\,\omega_T
-\sum_{n=0}^N B_n \left( \sqrt{\epsilon}\right)^{n+1} \Big\lvert = 0.
\end{equation}
\end{proposition}

\proof
The solution, $\hat{z}$, of~\eqref{eq:implicitfthe} given by the implicit function Theorem is
analytic w.r.t $x\in\mathbb{T}$ and $\mu$ in a neighbourhood of $0$, $\frac{\partial H^*}{\partial
z}\Big \rvert_{h_0}\!\!\!\!\left(x,0\right)$ is positive for $h_0>1$, then for $\mu$ close enough to
the origin $\frac{\partial H^*}{\partial z}\Big \rvert_{h_0}\!\!\!\!\left(x,\mu\right)$ is positive
and analytic w.r.t to $x\in \mathbb{T}$ and $\mu$. We can then write the following Taylor
development
\begin{equation}
  \left(\frac{\partial H^*}{\partial z}\Big
\rvert_{h_0}\!\!\!\!\left(x,\mu\right)\right)^{-1}=\sum_{n\geq 0}K_n\left(x,h_0\right)\mu^n,
\label{eq:defKn}
\end{equation}
where $\left(K_n\right)_{n\geq 0}$ are some known functions of $\left(x,h_0\right)$.
Using~\eqref{eq:develop} we get
\begin{equation}
  \label{eq:firstorderK}
  K_0\left(x,h_0\right)=\frac{1}{\sqrt{2\left(h_0-\cos x\right)}} \text{ and }
K_1\left(x,h_0\right)=0,
\end{equation}
we left to remark~\ref{rem:calculationK} an iterative schema to calculate $K_n$.

\indent
For all $k\geq 0$ we can integrate the $K_n\left( x,h_0\right)$ and we rewrite~\eqref{eq:integrale}
as
\begin{equation*}
  \frac{2\pi}{\omega_*}=2\pi\sum_{n\geq 0}C_n\left(h_0\right)\mu^n,
\end{equation*}
from~\eqref{eq:firstorderK} we get
\begin{equation*}
  C_0\left(h_0\right)=\frac{1}{\omega_{pend}\left(h_0\right)}=\int_0^{2\pi} \frac{dx}{\sqrt{2
\left(h_0 - \cos x \right)}} \text{ and } C_1\left( h_0\right)=0
\label{eq:defCn}
\end{equation*}
where $\omega_{pend}\left( h_0\right)$ is the frequency of a rotation orbit of energy $h_0$ of the
pendulum $H_p\left(x,z\right)=\frac{z^2}{2}+\cos x$.

\indent
Clearly for all positive integer $N$ we have
\begin{equation}
\lim_{\mu\rightarrow 0} \mu^{-N}  \Big\lvert
\frac{1}{\omega_*\left(h_0\right)}-\sum_{n=0}^NC_n\left(h_0\right)\mu^n\Big\rvert=0.
\label{eq:firstestim}
\end{equation}
We show now that we can obtain a similar estimate for $\omega_*$. Let us set, for all positive
integer $N$, $S_N\left(\mu,h_0\right)=\sum_{n=0}^NC_n\left(h_0\right)\mu^n$. Because
$S_N\left(0,h_0\right)\neq 0$ we can express $1/S_N$ as a $\mu$--power series
\begin{align}
  \frac{1}{S_n\left(\mu,h_0\right)}&=\frac{1}{C_0\left(h_0\right)}+\sum_{n\geq
2}\mu^n\sum_{m=\lfloor \frac n N\rfloor +1}^{\lfloor \frac n 2\rfloor}\frac{\left(
-1\right)^m}{C_0^{m+1}\left(h_0\right)}\sum_{\substack{n_1+\ldots+n_m=n \\ 2\leq n_i\leq
N}}C_{n_1}\left(h_0\right)\ldots C_{n_m}\left(h_0\right) \notag \\
            &=\sum_{n \geq 0}B_n\left( h_0\right)\mu^n,
\label{eq:secestim}
\end{align}
where the second line is the definition of the $B_n$'s, $B_0\left(
h_0\right)=C_0\left(h_0\right)^{-1}$ and $B_1\left( h_0\right)=0$. Recalling~\eqref{eq:firstestim}
and \eqref{eq:secestim} we can then conclude that for all positive integer $N$
\begin{equation*}
\Big\lvert \lvert \mu \rvert^{-N}\left(\omega_*-\sum_{n=0}^N\mu^n B_n\left( h_0\right)\right)-\lvert
\mu \rvert^{-N}\sum_{n\geq N+1}\mu^n B_n\left( h_0\right)\Big\rvert =\frac{\omega_*}{S_N\left(
\mu,h_0\right)}{\it o}\!\left( \lvert \mu\rvert^N\right)
\end{equation*}
and
\begin{equation*}
\lim_{\mu\rightarrow 0}\lvert \mu \rvert^{-N}  \Big\lvert \omega_*-\sum_{n=0}^N\mu^n B_n\left(
h_0\right)\Big\rvert =0.
\end{equation*}
\indent
From~\ref{eq:frequencies} and corollary~\ref{pro:analyticprop} we get for all positive integer $N$ 
\begin{equation*}
  \Big\lvert \omega_T -\sum_{n=0}^NB_n\left(h_0\right)\left(\sqrt{\epsilon}\right)^{n+1} \Big\rvert
\leq \epsilon De^{-\frac{D^{\prime}}{\sqrt{\epsilon}}}+\sqrt{\epsilon}{\it o}\!\left(
\sqrt{\epsilon}\right)^{N},
\end{equation*}
dividing by $\left(\sqrt{\epsilon}\right)^{N+1}$ and passing to the limit $\epsilon \rightarrow 0$
we obtain~\eqref{eq:asympexp}.
\endproof

From this Proposition Lemma~\ref{lem:lemma1} follows immediately.

\begin{remark}
\label{rem:calculationK}
The calculation of the $K_n$'s defined by~\eqref{eq:defKn} is a little cumbersome but easy, it needs
only some algebraic manipulations of power series. We assume known the solution, $\hat{z}\left( x,
\mu,h_0\right)$, of~\eqref{eq:implicitfthe} given by the implicit function Theorem, and using the
analyticity hypothesis we write it as a $\mu$--power series
\begin{equation*}
  \hat{z}\left(x,\mu,h_0\right)=\hat{z}_0\left( x,h_0\right)+\sum_{n\geq 1}\hat{z}_n\left(
x,h_0\right)\mu^n.
\end{equation*}

\indent
Starting from $H^*$ given by~\eqref{eq:firstterms}, we calculate its partial derivative w.r.t. $z$,
then we substitute $z=\hat{z}$ and we reorder the powers of $\mu$ to get
\begin{align*}
  \frac{\partial H^*}{\partial
z}\left(x,\hat{z},\mu\right)&=\hat{z}_0\left(x,h_0\right)+\sum_{l\geq2}\hat{z}_l\mu^l+\sum_{l\geq
3}\mu^l\sum_{\substack{n+m=l\\m\geq 1,2\leq n\leq
N}}\sum_{k=1}^m\frac{1}{k!}\frac{\partial^{k+1}H_n}{\partial z}\left(x,\hat{z}_0\right)\cdot \\
&\sum_{\substack{m_1+\ldots+m_k=m\\ m_i\geq1}}\hat{z}_{m_1}\ldots
\hat{z}_{m_k}=\hat{z}_0\left(x,h_0\right)+\sum_{l\geq 2} \mathcal{H}_l\left(x,h_0\right) \mu^l.
\end{align*}
This series is invertible in the field of the formal $\mu$--power series, then~\eqref{eq:defKn}
defines correctly a $\mu$--power series, which is given by
\begin{align*}
\left(\frac{\partial H^*}{\partial
z}\left(x,\hat{z},\mu\right)\right)^{-1}&=\frac{1}{\hat{z}_0}+\sum_{l\geq 2}\mu^l
\sum_{m=1}^{\lfloor \frac l 2 \rfloor} \frac{\left(
-1\right)^m}{\hat{z}_0^{m+1}}\sum_{\substack{l_1+\ldots+l_m=l\\ l_i\geq 2}}\mathcal{H}_{l_1}\ldots
\mathcal{H}_{l_m}\\
  &=K_0\left(x,h_0\right)+\sum_{l\geq2}K_l\left(x,h_0\right) \mu^l.
\end{align*}
\end{remark}

\section{Numerical Analysis of the scaling law at the resonance $p /q$.}
\label{secnaff}

\subsection{Frequency Map Analysis.}

\indent
The Frequency Map Analysis  of Laskar (FMA) is a numerical method which allows to
obtain a global view of the dynamics of  Hamiltonian systems 
by studying the properties of the frequency map, numerically defined from the 
action like variables  to the frequency space  using adapted Fourier techniques.
Thanks to its
precision it was used for the study of stability questions and/or diffusion properties of
a large class of
dynamical systems : Solar System \cite{Laskar2}, Particles Accelerator
\cite{LaskarRobin}, Galactic dynamics \cite{PAP96,PAP98}, Standard Map \cite{Laskar1,LaskarFroeschleCelletti}.
We present here the outlines of the method, following \cite{Laskar3}.

\subsubsection{Frequency Maps}
Considering a $n$--degrees of freedom quasi--integrable Hamiltonian system in 
the form
\begin{equation}
  H\left(J,\theta;\epsilon\right)=H_0\left(J\right)+\epsilon H_1\left(J,\theta\right)
\label{eq:hamsys}
\end{equation}
where $H$ is real analytic for $\left(J,\theta\right)\in B\times \T^n$, being $B$ an open domain in
$\R^n$, and $\epsilon$ is a real ``small'' parameter.
For $\epsilon=0$ this system reduces to an integrable one.
The motion takes place on invariant tori $J_j=J_j(0)$ described at constant
velocity $\nu_j\left(J\right)=\frac{\partial H_0}{\partial J_j}\Big\rvert_{J\left(0\right)}$,
 for $j=1,\ldots,n$.
Assuming a non--degenerate condition on $H_0$ the  Frequency Map  $F:B\rightarrow \R^n$
\begin{equation*}
F:J\mapsto F\left(J\right)=\nu
\end{equation*}
is a diffeomorphism on its image $\Omega$. In this case KAM theory~\cite{Kolmogorov,Arnold,Moser}
insures that for sufficiently small values of $\epsilon$, there exists a Cantor set
$\Omega_{\epsilon}\subset \Omega$ of frequency vectors satisfying a diophantine condition
\begin{equation*}
  \lvert <k,\nu>\rvert >\frac{C_{\epsilon}}{\lvert k\rvert^m}
\end{equation*}
for some positive constants $C_{\epsilon}$ and $m$, for which the quasi--integrable
system~\eqref{eq:hamsys} still possess smooth invariant tori, $\epsilon$--close to the tori of the
umpertubed system, with linear flow $t\mapsto \nu_j t+\theta_j\left(0\right) \,\mod 2\pi$ for
$j=1,\ldots,n$. Moreover, according to P\"oschel~\cite{poschel} there exists a diffeomorphism
$\Psi:\T^n\times \Omega \rightarrow \T^n\times B$
\begin{equation*}
  \Psi:\left(\phi,\nu\right)\mapsto\left(\theta,J\right)
\end{equation*}
which is analytic with respect to $\phi$ in $\T^n$ and $\mathcal{C}^{\infty}$ w.r.t $\nu$ in
$\Omega_{\epsilon}$, and which transforms the Hamiltonian system~\eqref{eq:hamsys} into
\begin{equation*}
  \begin{cases}
   \frac{d\nu_j}{dt}\left(t\right)=0\\ 
   \frac{d\phi_j}{dt}\left(t\right)=\nu_j
  \end{cases}
\end{equation*}
For frequency vectors $\nu \in \Omega_{\epsilon}$ the invariant torus can be represented in the
complex variables $(z_j=J_je^{i\theta_j})_{j=1,n}$  by a quasiperiodic expression
\begin{equation}
  \label{eq:cmplxtorus}
  z_j\left(t\right)=z_j\left(0\right)e^{i\nu_jt}+\sum_{m}a_{j,m}\left(\nu\right)e^{i <m,\nu> t}
\end{equation}

\indent
If we take the section $\theta=\theta_0$, for some $\theta_0 \in \T^n$, of the phase space, we
obtain a frequency map $F_{\theta_0}:B\rightarrow \Omega$
\begin{equation}
  \label{eq:freqmap}
  F_{\theta_0}:J\mapsto \pi_2\left(\Psi^{-1}\left(\theta_0,J\right)\right) 
\end{equation}
where $\pi_2\left(\psi,\nu\right)=\nu$ is the projection on $\Omega$. For sufficiently small
$\epsilon$ the non--degeneracy condition ensure that $F_{\theta_0}$ is a smooth diffeomorphism.

\subsubsection{Quasiperiodic approximations.}

If we have the numerical values of a complex signal over a finite time span $\left[ -T, T\right]$ and we think
that it has a quasiperiodic structure represented by a quasiperiodic function $f \left( t
\right)$, we can construct a quasi periodic approximation, $f^\prime \left( t \right)$, as follows.
We represent the given signal as:
\begin{equation}
f \left( t \right) = e^{i \nu_1 t}+\sum_{k\in \mathbb{Z}^n \setminus \left( 1,0,\dots,0 \right)} a_k
e^{i <k, \nu >t}  \quad a_k \in \mathbb{C},
\label{signalrepr}
\end{equation}
and the signal reconstructed by FMA as $f^\prime \left( t \right)=\sum_{k=1}^N a^\prime_k e^{i
\omega^\prime_k t}$. Frequencies $\omega^\prime_k$ and amplitudes $a^\prime_k$ are determined with
an iterative scheme. 
We first determine the first frequency $\omega^\prime_1$ as the value of $\sigma$ which maximise:
\begin{equation*}
\phi \left( \sigma \right)=\Big\lvert \frac{1}{2T} \int_{-T}^T f \left( t \right) e^{-i \sigma t}
\chi\left( t \right) dt \Big\rvert,
\label{amplitude}
\end{equation*}
where $\chi\left( t \right)$ is some weight function\footnote{A weight function is a positive and
even function such that:
\begin{equation*}
\frac{1}{2T}\int_{-T}^T \chi \left( t \right) dt = 1.
\end{equation*}
In all the computations we will take the Hanning filter: $\chi_1 \left( t \right) = 1 + \cos
\frac{\pi t}{T}$.}.
 Once the first frequency is found the associated complex amplitude $a_1^\prime$ is obtained by
orthogonal projection. We then iterate the scheme to the function: $f_1\left( t \right) =f\left( t
\right) -a_1^\prime e^{i \omega_1^\prime t}$.

\indent
Given the signal (\ref{signalrepr}) over a time span $\left[ -T, T \right]$, if we assume some
arithmetic properties of the frequencies (Diophantine numbers for example), then one can prove
(\cite{Laskar3}) that, using the Hanning filter of order $p$: $\chi_p \left( t \right)=\frac{2^p
\left( p! \right)^2}{\left( 2p \right)!}\left( 1+\cos \pi t\right) ^p$, FMA converges towards the first
frequency $\nu_1^T$ with the asymptotic expression for $T \rightarrow +\infty$:
\begin{equation}
\nu - \nu_1^T = \frac{\left( -1 \right)^{p+1} \pi^{2p} \left( p! \right)^2}{A_p T^{2p+2}} \sum_{k}
\frac{\Re a_k}{\Omega_k^{2p+1}}\cos \left( \Omega_k T \right) + {\it o}\! \left(
\frac{1}{T^{2p+2}}\right),
\label{approxsol}
\end{equation}
where $\Omega_k = <k,\nu> - \nu_1$ and 
$A_p=-\frac{2}{\pi^2}\left(\frac{\pi^2}{6} - \sum_{k=1}^{p}
\frac{1}{k^2} \right)$.
We can thus note that the accuracy of the determination of the frequency 
is $O(1/T^{2p+2})$, which will usually be several
order of magnitude better that with usual FFT ( $O(1/T)$).

\subsubsection{Frequency Map Analysis}
With the previous algorithm, it is  possible to construct numerically a frequency map
in the following way : a) Fix all values of the angles $\theta_j=\theta_{j0}$. b) for any value 
$J_0$ of the action variables, integrate numerically the trajectories of initial condition 
$(J_0,\theta_0)$ over the time span $T$. c) Search for a quasiperiodic approximation of the trajectory 
with the algorithm of the previous section. d) Identify the fundamental frequencies $\nu$ of this 
quasiperiodic approximation. The frequency map is then

\begin{equation}
  \label{eq:freqmap1}
  F_{\theta_0}^T:J\mapsto \nu
\end{equation}
and from the previous section, we know that for $T\longrightarrow +\infty$, on the set 
of regular KAM solutions,  
$F_{\theta_0}^T \longrightarrow F_{\theta_0}$. In particular, $F_{\theta_0}^T$ should converge 
towards a smooth function of the set of invariant KAM curves.
The destruction of these invariant curves can thus be identified by the non regularity of the 
frequency map $F_{\theta_0}^T $.

\subsection{Application to the Standard Map.}
\label{applicsm}

We are interested in the study of homotopically non--trivial invariant orbits of the Standard map,
in particular in the relation between the rotation number $\omega$ and the critical function
$\epscrit$. Given a couple of initial conditions $\left( x_0,y_0 \right)$ the FMA gives us
approximate amplitudes and frequencies of the resulting orbit.
For $\epsilon =0$ all invariant curves are tranverse to section $x=constant$, for $\epsilon$
sufficiently small this fact still holds, that is the tranversality property.
For $\epsilon >0$, ``curves'' of rational frequency do not persit and are replaced by
``set of periodic fixed points'' such that if $\omega \left( x_0,y_0
\right)=p/q$ then this set contains $2q$ periodic points, an half of which are hyperbolic (HFP) and
the other elliptic (EFP).

\begin{center}
  \begin{figure}[ht] 
   \makebox{
\includegraphics[scale=0.5]{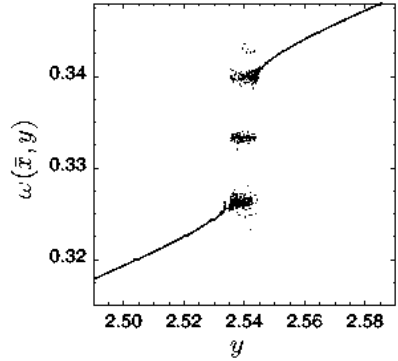}     
    }
  \caption{Particular of a nearby zone of a HFP, section $\bar{x}=4.408$ and $\epsilon=0.7$
(neighbourhood of the $1/3$ resonance). Note that for $y\in \left[2.535,2.545 \right]$ the frequency
curve looks very irregular.}
  \label{jumps}
  \end{figure}
\end{center}

\indent
Near a HFP appears a stochastic layer, that is there are not invariant curves with frequency ``very
close'' to the frequency of the HFP. This is shown in Figure~\ref{jumps} where 
the stochactic layer is revealed by the nonregularity of the frequency map around
the frequency value $1/3$. 

For a fixed value of $\epsilon$ and a HFP of rotation number $p/q$, we study the 
frequency curve $y \mapsto \omega
\left(x,y\right)$ for $y$ in some neighbourhood, $\mathcal{I}$, of $y_{H\!F\!P}$, and
$x=x_{H\!F\!P}$ fixed. We generalise in an obvious way the definitions~\eqref{def:omegap} and \eqref{def:omegam}
of $\omega^+(\epsilon)$ and  $\omega^-(\epsilon)$ defined in the vicinity of rotation number $0$
 for  invariant curves in the vicinity 
of a general HPF of rotation number $p/q$ as  $\omega_{p/q}^+(\epsilon)$, $\omega_{p/q}^-(\epsilon)$ 
and $\Delta \omega_{p/q}= \omega^{+}_{p/q} - \omega^{-}_{p/q}$. Similarly we extend the definitions
of $y^{\pm}\left( \epsilon\right)$ as $y^{\pm}_{p/q}\left(
\epsilon\right)$.

Numerically
$y^{+}_{p/q}\left( \epsilon\right)$ is determined as the smallest value of $y(0)$, 
larger than $y_{H\!F\!P}$, for which the frequency curve is still regular (see
appendix~\ref{numericalhint}), and similarly
for $y^{-}_{p/q}\left( \epsilon\right)$. Then we assume $\omega^{\pm}_{p/q}\left( \epsilon\right)=
\omega \left( x_{H\!F\!P}, y^{\pm}_{p/q}\left( \epsilon\right) \right)$.
Studying $\Delta \omega_{p/q}\left(\epsilon\right)$ as a function of $\epsilon$ we are then able to
reconstruct the curve $\epsilon_{crit}\left( \Delta\omega_{p/q} \right)$ for a fixed resonance. In
fact the non-existence of invariant curves, for a fixed value of $\epsilon$, with frequency
$\tilde{\omega}$, between $\omega^{+}_{p/q}$ and $\omega^{-}_{p/q}$ means that
$\epsilon_{crit}\left( \tilde{\omega} \right) > \epsilon$ and $\epsilon_{crit}\left( \omega_{p/
q}^{\pm} \right)=\epsilon$.

\subsection{Presentation of the numerical results.}
\label{result}
\begin{center}
  \begin{figure}[ht]
   \makebox{
\includegraphics[scale=0.5]{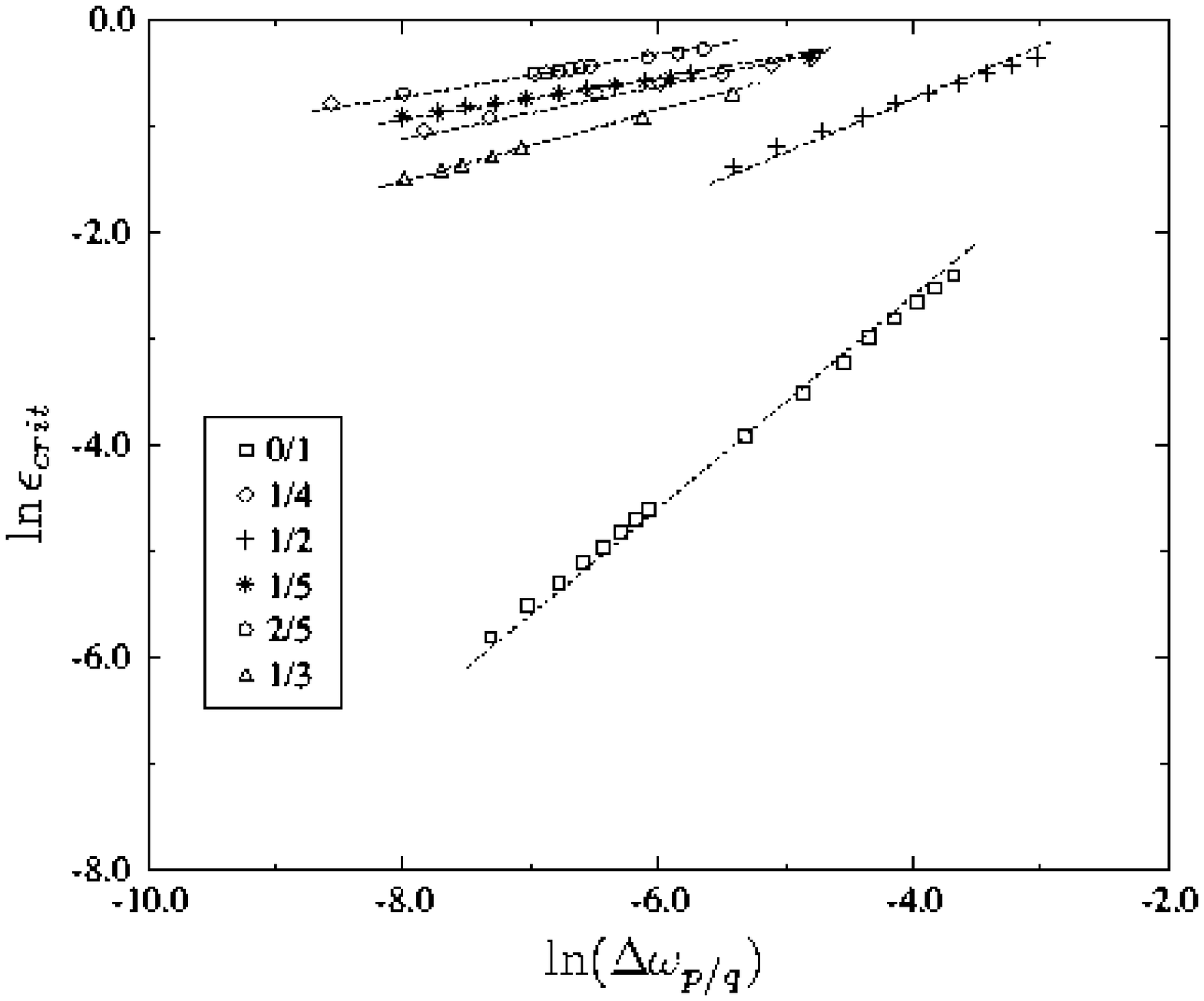} 
    }
  \caption{Plot of $\ln \epsilon_{crit}$ vs $\ln \left( \Delta \omega_{p/q} \right)$ for $p/q \in
\{0/1,1/5,1/4,2/5,1/3,1/2 \}$.}
  \label{res01fig}
  \end{figure}
\end{center}

Using the FMA we studied the behaviour of $\epsilon_{crit}\left( \Delta \omega_{p/q}\right)$ in
neighbourhood of the resonances $\frac 0 1,\frac 1 5,\frac 1 4 ,\frac 2 5,\frac 1 3,\frac 1 2$.
The results are presented in Table~\ref{tableres} and in Figure~\ref{res01fig}, where the plot of $\ln
\epsilon_{crit} \left( \Delta \omega_{p/q} \right)$ as a function of $\ln \Delta \omega_{p/q}$ is
shown. All numerical calculations are done with $100$ digits 
accuracy using the Fortran $90$ multi--precision  package {\em
mpfun}~\cite{Bailey}. 
\begin{table}[ht]
  \begin{center}
    \begin{tabular}[ht]{|c|c|c|}
\hline \hline
resonance $p/q$ & $A_{p/q}$ & $C_{p/q}$ \\
\hline \hline
 $0/1$          & $0.937$       & $1.066$ \\     
\hline
 $1/5$          & $0.175$       & $0.489$ \\     
\hline
 $1/4$          & $0.223$       & $0.723$ \\     
\hline
 $2/5$          & $0.175$       & $0.714$ \\     
\hline
 $1/3$          & $0.312$       & $1.000$ \\     
\hline
 $1/2$          & $0.427$       & $0.957 $ \\ 
\hline \hline
    \end{tabular}
 \smallskip    
    \caption{$\ln \epsilon_{crit} = A_{p/q} \ln \Delta \omega_{p/q} + C_{p/q}$, 
    where the coeffcients $A_{p/q}$ and $C_{p/q} $ are determined by 
    a linear least square fit to the FMA results.}
    \label{tableres}
  \end{center}
\end{table}

\indent
These results suggest the following geometry  of the $\epsilon_{crit}$--$\omega$ plane. Consider only
the resonance $p/q$, then for $\omega$ close to $p/q$ we have:
\begin{equation}
  \label{oneres}
  \epscrit \sim C^{\prime}_{p/ q} \Big \lvert \omega - \frac p q \Big \rvert^{\frac 1 q}
\end{equation}
Because of the density of the rational numbers on the real line, close to $p/q$ there are infinitely
many rational numbers $p^{\prime}/q^{\prime}$, with $q^{\prime}$ bigger
and bigger as $p^{\prime}/q^{\prime}$ is closer to $p / q$.
The relation between $\epsilon_{crit}$ and $\omega$ is not the simple one given by \eqref{oneres},
even in a very small neighbourhood of $p/q$, but we must add the contributions of all
$p^{\prime}/q^{\prime}$ near $p /q$. These contributions are nevertheless negligible because for
very big $q^{\prime}$ the right hand side of \eqref{oneres} is almost equal to one. Then the local
description given by the \eqref{oneres} should be a very good approximation of the relation between
$\epsilon_{crit}$ and $\omega$ (see Figure~\ref{resondom}).
\begin{center}
  \begin{figure}[ht]
   \makebox{
\unitlength=1.000000pt
\begin{picture}(230.00,140.00)(0.00,0.00)
\put(70.00,10.00){\makebox(0.00,0.00){$\frac 1 4$}}
\qbezier(70.00,20.00)(70.00,80.00)(50.00,90.00)
\qbezier(70.00,20.00)(70,80.00)(90.00,90.00)
\put(100.00,10.00){\makebox(0.00,0.00){$\frac 1 3$}}
\qbezier(100.00,20.00)(90.00,80.00)(70.00,90.00)
\qbezier(100.00,20.00)(110.00,80.00)(130.00,90.00)
\qbezier(180.00,20.00)(180.00,100.00)(120.00,110.00)
\put(180.00,10.00){\makebox(0.00,0.00){$\frac 1 2$}}
\put(20.00,20.00){\line(4,5){80.00}}
\put(30.00,10.00){\makebox(0.00,0.00){$0$}}
\put(0.00,130.00){\makebox(0.00,0.00){$\epsilon_{crit}$}}
\put(220.00,0.00){\makebox(0.00,0.00){$\omega$}}
\put(0.00,20.00){\vector(1,0){230.00}}
\put(19.00,10.00){\vector(0,1){130.00}}
\end{picture}
    }  
    \caption{The resonant domains in the $\epsilon_{crit}$--$\omega$ plane.}
    \label{resondom}
  \end{figure}
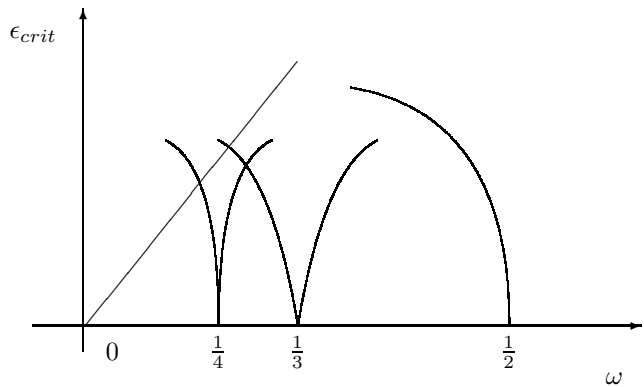
\end{center}

\section{Conclusions.}

We have studied both analytically and numerically the Standard Map critical function in a neighbourhood
of a fixed resonance. For the fundamental resonance we find a linear scaling law, whereas
in~\cite{BerrettiGentile1} and \cite{BerrettiGentile2} a quadratic scaling law was found. We think
these two results agree, in fact we think there is a difference in the scaling law for real or
complex frequencies. This shows in particular that for diophantine values of the rotation number 
$\omega$, sufficiently close to zero, the critical function $\epsilon_{crit}(\omega)$ 
and the critical radius of convergence  $\rho_{crit}(\omega)$ are different. More precisely
$\epsilon_{crit}(\omega) >  \rho_{crit}(\omega) > 0 $ 

We stress again that this result implies that the conjugating function $u$ defined as the sum of the 
Standard Map Lindstedt series \eqref{Lindstedt} for $\lvert\epsilon\rvert < \rho_{crit}(\omega)$, can be 
analytically continued outside its disk of convergence at least for $\epsilon$ real. This implies that
in the plane $\epsilon$--complex the domain of convergence of the Lindstedt series is not a disk; it will
 be interesting to study its geometry, in particular its boundary: it could be a fractal curve as in the 
 case of the Schr\"oder--Siegel center problem \cite{CarlesonGamelin}, we only know that the nearest singularity
 to the origin is on the imaginary axis.

\indent

\indent
The proof of Theorem~\ref{the:maintheorem} is different from the one of~\cite{BerrettiGentile1}.
The FMA method doesn't allow us (at least for the moment) to use complex frequencies so we couldn't
verify the results of~\cite{BerrettiGentile1} directly.
 It
would be interesting to investigate the scaling law for the generic resonance $p/q$ with our
methods, but the proof of Theorem~\ref{the:maintheorem} is specific to the fundamental resonance: we
don't have a bound of the initial data for the last invariant torus for a generic resonance as we
proved in Proposition~\ref{pro:initialdata} for the resonance $0/1$.
Supported by the numerical results of section \ref{secnaff}, we can 
nevertheless make the following conjecture
\begin{conjecture}
For all rational $p/q$, we can find a positive constant $c_{p/q}$, such that for
all $\omega$ in a small {\em real neighbourhood} of $p/q$
\begin{equation}
\epscrit \sim c_{p/ q} \Big\lvert \omega - \frac p q \Big\rvert ^{\frac 1 q}.
\label{eq:conjecture}
\end{equation} 
\end{conjecture}

\indent
We can make a link with the Brjuno function. The remark in footnote~\ref{ftn:scallawbrjuno}
with Theorem~\ref{the:maintheorem} imply that there exists a
neighbourhood of the origin where $B\left( \omega \right) + \ln\epscrit$ is uniformly bounded. If we
are able to prove the previous conjecture the same bound will hold in the neighbourhood of every
rational, that is we could prove $B\left( \omega \right) + \ln\epscrit$ is uniformly bounded for all
frequencies.

\indent
In this paper we were interested in finding the optimal scaling law, namely the optimality of the
exponent $1/q$ in~\eqref{eq:conjecture}, neglecting the behaviour of the proportionality constants
$c_{p/q}$ as function of $p/q$. It will be interesting to investigate (at least numerically) this
function, in particular to test its continuity, and to search if the function
${\epsilon_{crit}\left( \Delta \omega\right)}/{\Delta \omega ^{\frac 1 q}}$ is continuous for all
rational numbers.

\appendix

\section{Numerical precision of the results.}
\label{numericalhint}

Here we analyse more precisely the numerical results, pointing out some important remarks
concerning the precision of the calculations.

The destruction of invariant curve is identified by the nonregularity of the frequency map.
In general, this nonregularity is clearly visible (fig 3), and 
the destruction of invariant curves is determined  with  no ambiguity.
But when one searches for very fine details, which correspond to small 
values of the perturbing coefficient $\epsilon$, 
one needs to be able to distinguish between nonregularity of the frequency map and 
the apparent nonregularity resulting from numerical errors in the computations.

\begin{center}
  \begin{figure}[ht]
   \makebox{
   \includegraphics[scale=0.55]{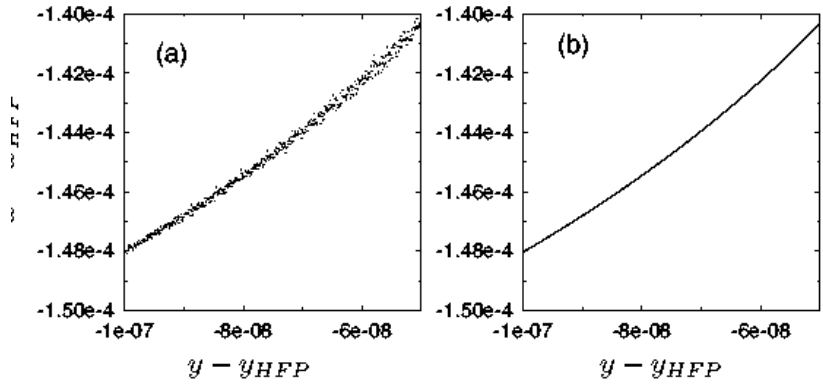}  
    }
   \caption{Frequency curve calculated in double precision  (a) and quadruple precision (b). Section
$x=4.408$ for the Standard Map for $\epsilon~=~0.1$, integration time $T=90000$. The regular set
$\mathcal{I}_{reg}$ is represented.}
   \label{figuresix}
  \end{figure}
\end{center}

One way to work around this problem is to search always for a regular part of the frequency 
curve. The existence of this regular part is the indication that 
we have not yet reached the  threshold  for numerical roundoff error. An illustration is 
given in figure \ref{figuresix}, where  the frequency curve is plotted 
for $\epsilon =0.1, T=90 000$, for both double and quadruple precision computations.
It is clear that with double precision computations, we are no longer able to distinguish 
precisely between regular and non regular behavior of the frequency curve, 
while it is still possible  with quadruple precision artihmetics.

As we know that on the regular  curves, the FMA algorithm converges towards the 
true frequency as $1/T^4$ with the Hanning window of order 1 ($p=1$  in Eq.\ref{approxsol}), we 
can check the errors of the FMA method by increasing $T$. If the variation of 
frequencies increases instead of decreasing, and still present a regular behavior 
as in Figure \ref{figuresix}.a, it should come from numerical noise, and needs to be checked 
further.
\begin{center}
  \begin{figure}[ht]
   \makebox{
   \includegraphics[scale=0.55]{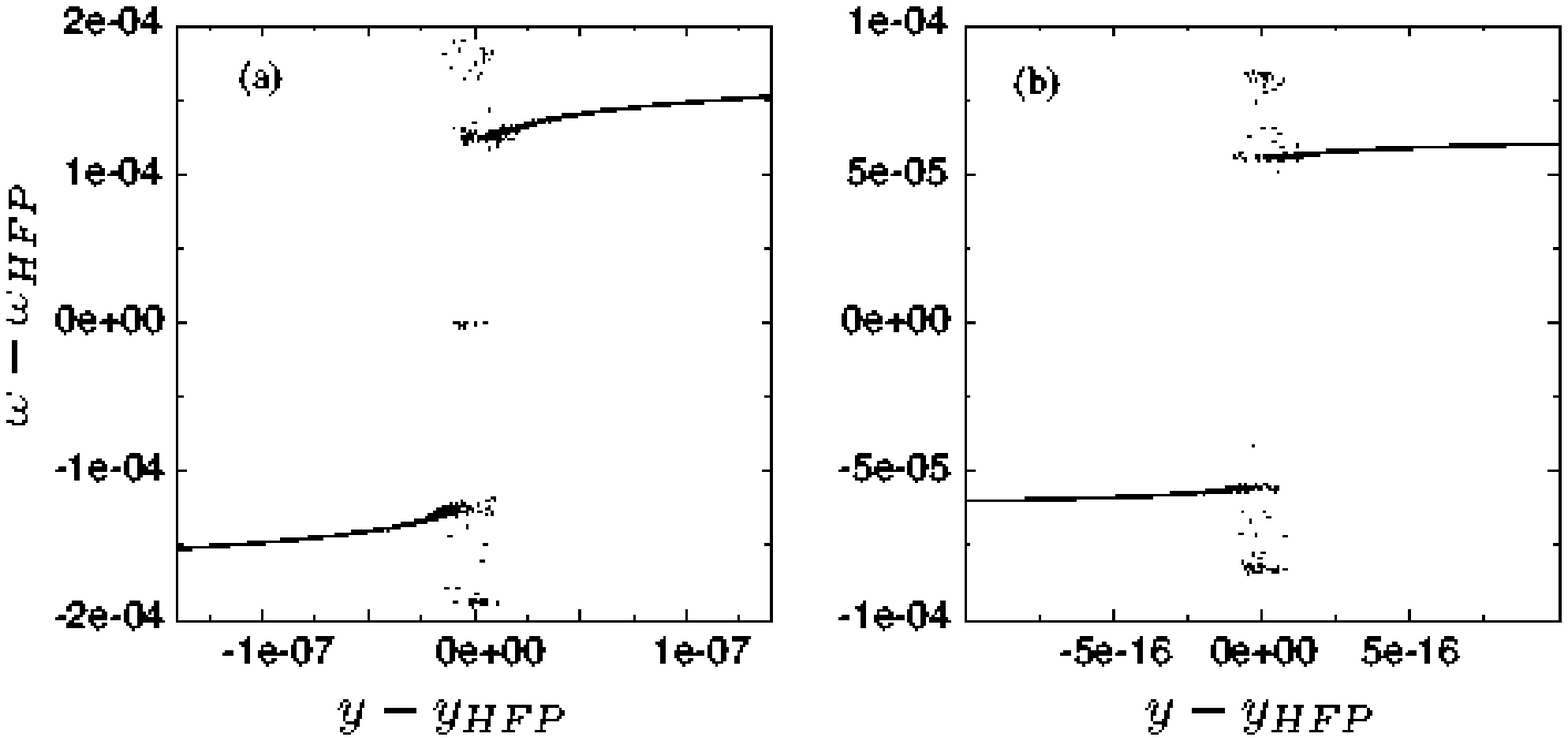}  
    }
   \caption{Double precision calculations (a) and Quadruple precision calculations (b) for the
section $x=x_{H\!F\!P}$ for the Standard map with $\epsilon=0.1$.}
   \label{figuresept}
  \end{figure}
\end{center}
%


\subsection{False interpretation of a first result.}
\label{interpretation}
In fact, the numerical accuracy for the iteration of the map is very important when 
one searches for very small values of $\epsilon$, and  not taking  it properly
into account can lead to false results.
Indeed, in
figures \ref{figuresept}, we show sections $x=x_{H\!F\!P}$ of the Standard
Map with $\epsilon=0.1$ for double precision (Fig. \ref{figuresept}.a) 
and quadruple precision (Fig. \ref{figuresept}.b) calculation. One can note a very large 
change in the $y$--range, which
 implies that  the values of $y^{-}_{p/q}$ and $y^{+}_{p/q}$ 
obtained using double precision are largely 
over--estimated.  In this case, we must use at least the quadruple
calculation, while for the final results of the present paper, 
we had to use  the $100$--digits multi-precision 
using the Fortran $90$  package  {\em mpfun}.

Because of the finite precision arithmetic there is a threshold $x_{thr}$ such that for all $x$
below the threshold then  $\sin x = x$. When we start with initial condition $\left( x_0,y_0
\right)= {\it o}\! \left( x_{thr} \right)$ then instead of iterating the standard
map~\eqref{standardmap} we iterate the linear map $T_{\epsilon}^{lin}$
\begin{align*}
T_{\epsilon}^{lin}: \begin{cases}
  y^{\prime} &= y + \epsilon x \\
  x^{\prime} &= x + y^{\prime} \quad \mod 2\pi .
                    \end{cases}
\end{align*}

\begin{center}
  \begin{figure}[ht]
   \makebox{
\includegraphics[scale=0.5]{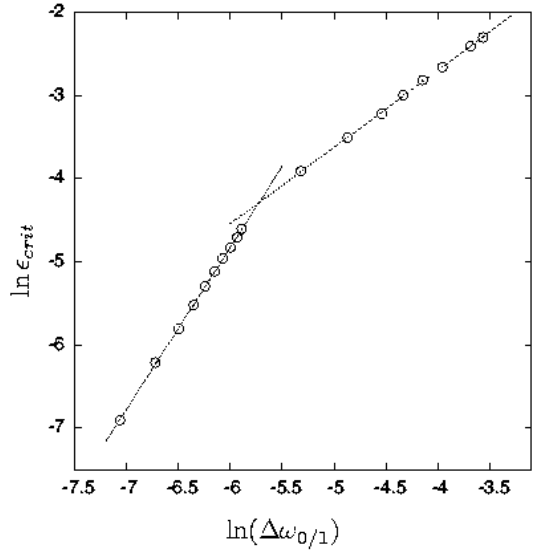} 
    }
  \caption{Change of slope phenomenon for the resonance $0/1$. $\ln
\epsilon_{crit}$ is plotted versus  $\ln \Delta \omega_{0/1}$ using  quadruple precision. The
linear fit  $\ln \epsilon_{crit}=1.948 \ln \Delta \omega_{0/1}+6.863$ and $\ln
\epsilon_{crit}=0.92 \ln \Delta \omega_{0/1}+0.998$ are also plotted.}
  \label{chpnt01}
  \end{figure}
\end{center}

\indent
The iteration of $T_{\epsilon}^{lin}$ produces a strange result which is misleading. Using data
below the threshold value $x_{thr}$, the iteration of $T_{\epsilon}^{lin}$ and then the numerical
analysis of the signal gives the following scaling law:
  There exists $0 < \epsilon^* <1$ such that:
  \begin{enumerate}
  \item if $\epscrit > \epsilon^*$ then: $\ln \epsilon_{crit}\left(\Delta
\omega_{\frac{p}{q}}\right)  \sim \frac{1}{q} \ln \Delta \omega_{\frac{p}{q}}+c_1$,
  \item if $\epscrit < \epsilon^*$ then: $\ln \epsilon_{crit}\left(\Delta
\omega_{\frac{p}{q}}\right) \sim \frac{2}{q} \ln \Delta \omega_{\frac{p}{q}}+c_2$;
  \end{enumerate}
 namely it seems to be a discontinuity in the scaling law as Figure~\ref{chpnt01} shows.
\indent
Looking at~\cite{BerrettiGentile1} and comparing with the results of~\cite{MarmiStark} at first
time, using the double precision, we interpreted the result of Figure~\ref{chpnt01} as follows:
\begin{quotation}{\it 
  The scaling law at resonances for the Standard Map is the one presented in
\cite{BerrettiGentile1}. The different result of \cite{MarmiStark} can be justified saying that they
didn't go deep enough inside the resonant region and the non linearity of the model added some
distortion . We are in presence of  two behaviours: close to the resonant
region we find the proper law, far away the distorted one. }
\end{quotation}

\begin{center}
  \begin{figure}[ht]
   \makebox{
\includegraphics[scale=0.5]{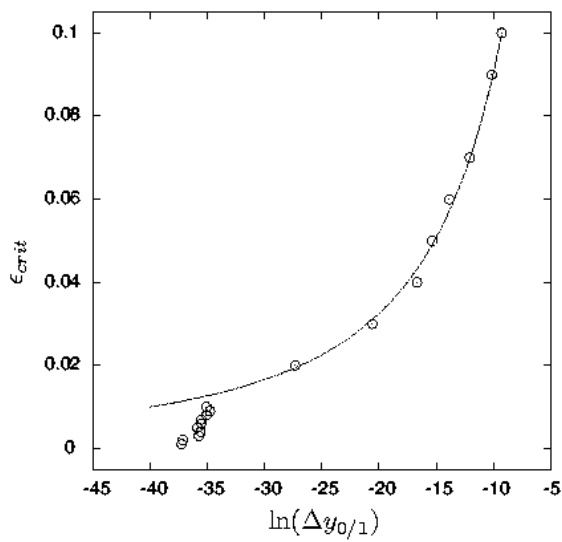} 
    }
  \caption{$\ln \Delta y_{0/1}$ as function of $\epsilon_{crit}$ for the fundamental resonance using
the quadruple precision. It is evident that for $\epsilon \leq 0.01$ then $\ln \Delta y_{0/1} \sim
-35$.}
  \label{deltayvse}
  \end{figure}
\end{center}

\indent
In studying further the reasons for this transition, we convinced ourselves that this interpretation
was not correct. We understood that this behaviour was due to the finite precision arithmetic
of  the iteration of the
map. We also observed that for  values of $\epsilon$ smaller than a critical value $\epsilon_k$,
there was a saturation of  the measured  value of $\Delta y_{0/1} =
y^{+}\left( \epsilon\right)-y^{-}\left( \epsilon\right)$, which was also the sign
of numerical problems (Fig.~\ref{deltayvse}). 
This was confirmed by the analytic result of Theorem~\ref{the:maintheorem} for the
resonance $0 / 1$.
\subsection{New interpretation.}

With Lemma~\ref{lem:lemma1} we proved that the rotation number of the standard map is very well
approximated by the pendulum frequency. Rewriting~\eqref{eq:quiteend} as follows
\begin{equation}
  \ln \epsilon_{crit} \sim 2 \ln \Delta \omega_{0/ 1} +C+ 2 \ln \ln
\frac{8\sqrt{\epsilon_{crit}}}{\Delta y_{0/1}}
\label{chpnt2}
\end{equation}
and using the previous remark on $\Delta y_{0/1}$, it follows that the term $\ln \ln
\frac{8\sqrt{\epsilon_{crit}}}{\Delta y_{0/1}}$ in \eqref{chpnt2} is almost constant
in the interval of variation of $\epsilon$, say for example $\epsilon \in [0.001,0.01]$. A similar
result holds for every resonance $p/q$; then for a fixed numerical precision we obtain the following
scaling law
\begin{equation*}
\ln \epsilon_{crit}\sim \frac 2 q \ln \Delta \omega_{p/ q}+c  
\end{equation*}

\begin{center}
  \begin{figure}[ht]
   \makebox{
\includegraphics[scale=0.55]{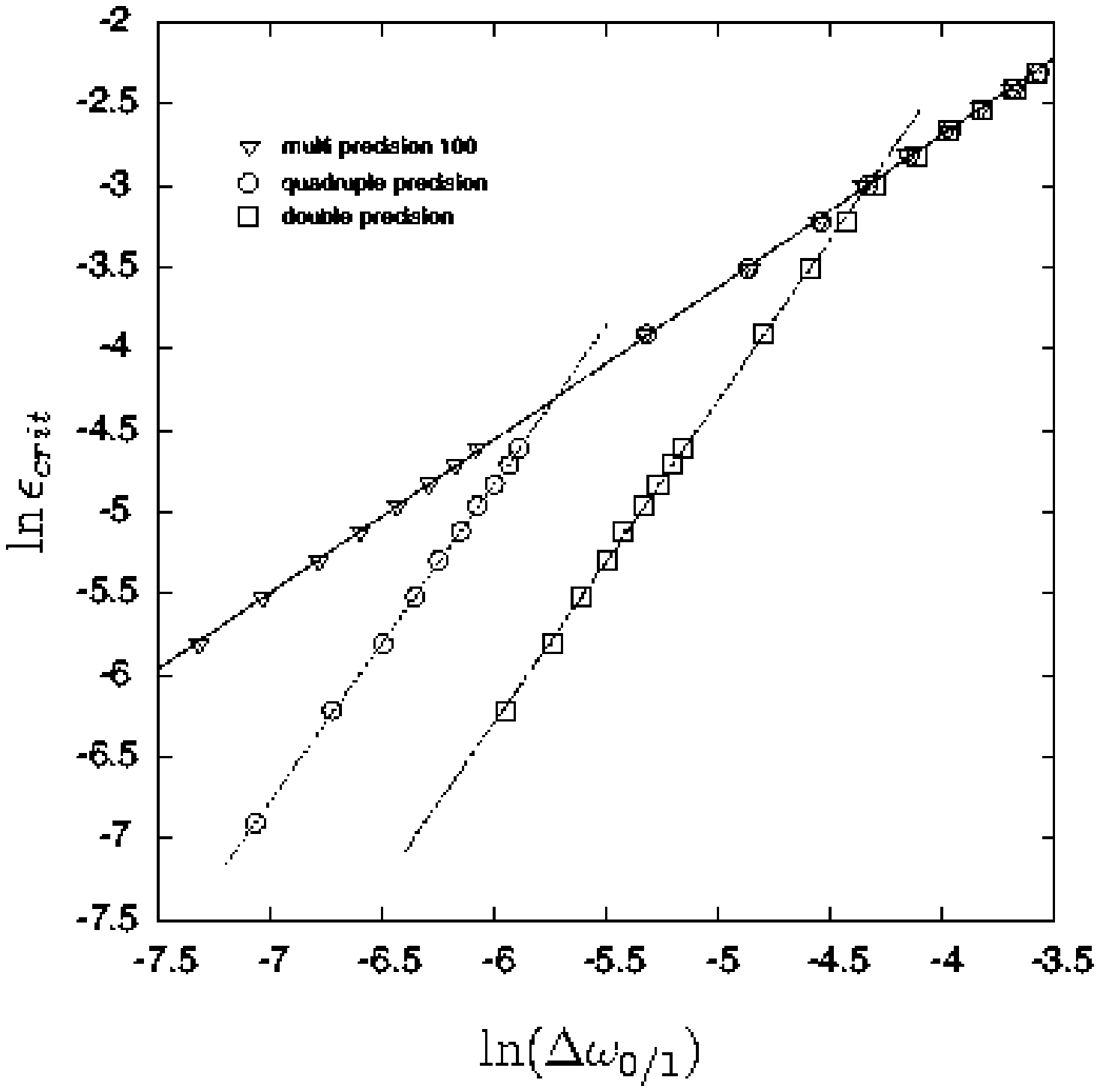} 
    }
  \caption{Plot of $\ln \epsilon_{crit}$ as a function of $\Delta \omega_{0/1}$ using the multiple
precision $100$, quadruple precision and double precision.}
  \label{alltog}
  \end{figure}
\end{center}

\indent
To prove that this change of slope is ``not real'' we increased the precision of the calculations
(i.e. passing from double precision to quadruple and then to multi-precision--$100$), obtaining that
the threshold $\epsilon^*$ decreases (see Figure~\ref{alltog}), showing that in the limit of exact
arithmetic this threshold doesn't exist.

\begin{center}
  \begin{figure}[ht]
   \makebox{
\includegraphics[scale=0.8]{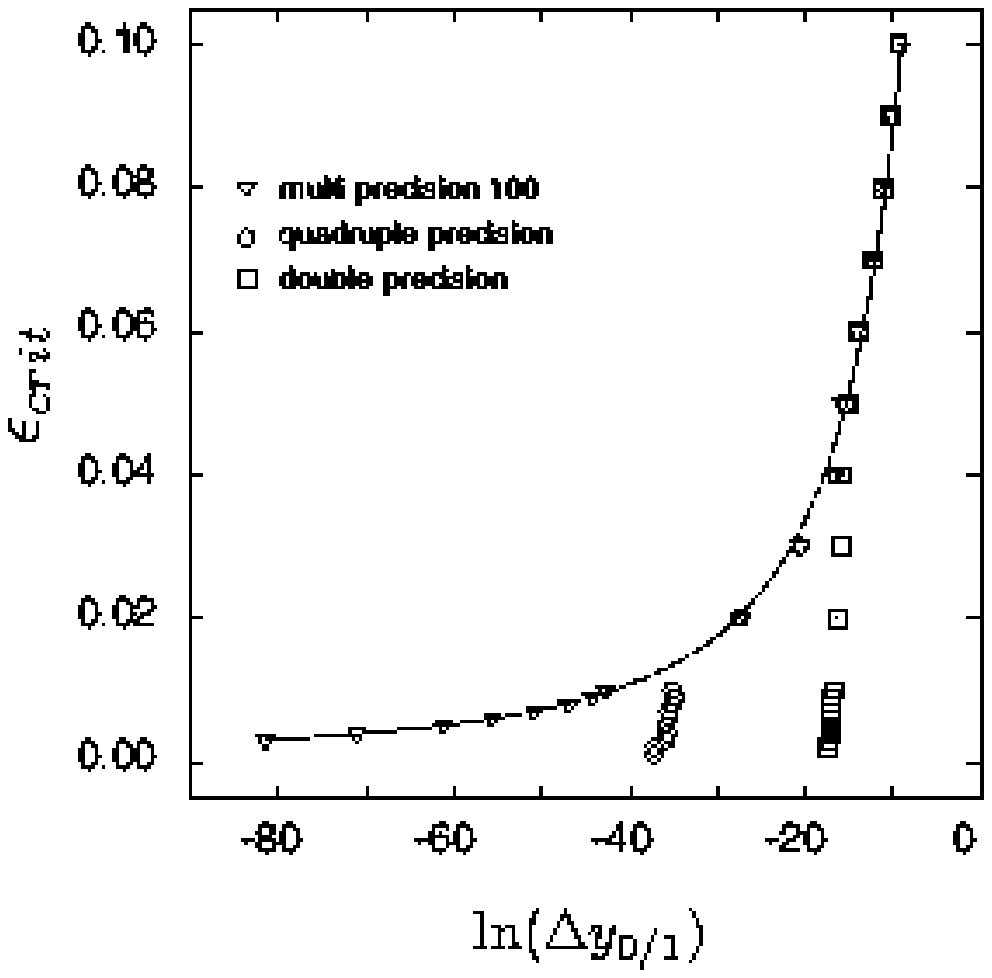}   
    }
  \caption{Plot of $\epsilon_{crit}$ vs $\ln \Delta y_{0/1}$ for different precisions. The solid
curve is the fitting of the multiprecision--$100$ data: $\ln \Delta
y_{0/1}~=~a+\frac{b}{\sqrt{\epsilon_{crit}}}$, with $a=6.177$ and $b=-4.804$. To compare with the
analytical result of Proposition~\ref{pro:initialdata} which gives $\ln \Delta y_{0/1} \geq
C-\frac{\pi^2}{2\sqrt{\epsilon_{crit}}}$. We don't have a good estimation for $C$ but we can compare
$b$ with $-\pi^2/2\sim-4.935$, exhibiting a good agreement between the numerical results and the
analytical one.}
  \label{logyvsefig}
  \end{figure}
\end{center}

\indent
Figure~\ref{logyvsefig} shows $\Delta y_{0/1}$ as a function of $\epsilon_{crit}$ for different
machine precision. Also in this case it is clear that the threshold value under which $y$ is
nearly constant goes to zero when we increase the numerical precision.

\indent
For the quadruple precision the threshold is $x_{thr} \sim 10^{-12}$ and it goes down to order
$10^{-32}$ using the multi-precision package $100$--digits. This value is sufficiently small to do
precise calculations in reasonable CPU--times. In fact the threshold value for the multi-precision
package $1000$--digits is order $10^{-333}$ but the CPU--time for the calculations increases several
order of magnitude, and this was not necessary to get good confidence with our present results.



\begin{thebibliography}{XX}


\bibitem{abramowitzstegun} M.Abramowitz and I.A.Stegun: {\it Handbook of mathematical functions with
formulas, graphics and mathematical tables}, Ed. Dover Pub., INC New York, $\left( 1972 \right)$.

\bibitem{Arnold} V.I. Arnold: {\it Proof of a Theorem of A.N.Kolmogorov on the invariance of
quasiperiodic motions under small perturbations of the Hamiltonian}, Usp. Mat. Nauk. $\mathbf{18}$
$13$, $\left( 1963 \right)$; Russ. Math. Surv., $\mathbf{18}$, $\left( 1963 \right)$, pp.$9-36$.

\bibitem{Bailey} D.H.Bailey: {\it A Fortran--$90$ based multiprecision system}, RNR Technical report
RNR--$94-013$, $\left( 1994 \right)$, {\it http://www.nas.nasa.gov/RNR/software.html}.

\bibitem{BenettinGiorgilli} G.Benettin, A.Giorgilli: {\it On the Hamiltonian interpolation of
near--to--the--identity symplectic mappings with application to symplectic integration algorithms},
Journal of Statistical Physics, Vol. $74$, Nos. $5/6$, $\left( 1994 \right)$.

\bibitem{BCCF} A.Berretti, A.Celletti, L.Chierchia and C.Falcolini: {\it  Natural boundaries for
area--preserving twist map}, J.Stat.Phys. $\mathbf{66}$, $\left( 1992 \right)$, pp.$1613$.

\bibitem{berrettichierchia1} A.Berretti and L.Chierchia: {\it  On the complex analytic structure of
the golden invariant curve for the standard map}, Nonlinearity $\mathbf{3}$, $\left( 1990 \right)$,
pp.$39-44$.

\bibitem{BerrettiGentile1} A.Berretti and G.Gentile: {\it Scaling properties for the radius of
convergence of a Lindstedt series: the Standard Map}, J.Math.Pures.Appl. $(9)$, $\mathbf{78}$,
$(1999)$, no.$2$, pp.$159-176$.

\bibitem{BerrettiGentile2} A.Berretti and G.Gentile: {\it Bryno function and the Standard Map},
preprint $(1998)$.

\bibitem{BerrettiMarmi} A.Berretti and S.Marmi: {\it Scaling near resonances and complex rotation
numbers for the standard map}, Nonlinearity $\mathbf{7}$, $\left(1994\right)$ $603-621$.

\bibitem{birkoff1} G.D.Birkhoff: {\it Proof of Poincar\'e's geometric theorem}, Trans. Am.Math.Soc.,
$\mathbf{14}$, $\left( 1913\right)$, pp. $14-22$.

\bibitem{birkoff2} G.D.Birkhoff: {\it An extension of Poincar\'e's last geometric theorem}, Acta
Math., $\mathbf{47}$, $\left( 1925\right)$, pp. $297-311$.

\bibitem{Brjuno} A.D.Brjuno: {\it Analytical form of differential equation}, Trans. Moscow Math.
Soc. $\mathbf{25}$, $\left( 1971 \right)$, pp.$131-288$.

\bibitem{CarlesonGamelin} L.Carleson and T.W.Gamelin: {\it Complex dynamics}, Springer--Verlag
$\left( 1993 \right)$.

\bibitem{carletti} T.Carletti: {\it Stability of orbits and arithmetics for some discrete dynamical
systems}, PhD Thesis Dep.Math.Univ.Florence, $(1999)$.

\bibitem{Chirikov1} B.V.Chirikov: {\it An universal instability of many dimensional oscillator
systems}, Phys.Rep. $\mathbf{52}$, $\left( 1979 \right)$, pp.$264-379$.

\bibitem{Davie1} A.M.Davie: {\it The critical function for the semi-standard map}. Nonlinearity
$\mathbf{7}$, $\left( 1990 \right)$, pp.$21 - 37$.

\bibitem{Davie2} A.M.Davie: {\it Renormalisation for analytic area--preserving maps}, University of
Edinburgh preprint $\left( 1995\right)$.

\bibitem{DumasLaskar} H.S.Dumas and J.Laskar: {\it Global dynamics and long--time stability in
Hamiltonian systems via numerical frequency analysis}, Phys.Rev.Lett., $\mathbf{7}$, $\left( 1993
\right)$, pp.$2975-2979$.

\bibitem{Gelfreich} V.G.Gelfreich: {\it A proof of the exponentially small transveraslity of the
separatrices of the standard map}, Communications in Mathematical Physics $\mathbf{201}$, $\left(
1999\right)$, pp. $155-216$.

\bibitem{Godement} R.Godement: {\it Analyse math\'ematique I}, $1^{st}$ edition, Springer--Verlag
Berlin $\left( 1998\right)$.

\bibitem{Greene1} J.M.Greene: {\it A method for determining a stochastic transition}, J.Math. Phys.
$\mathbf{20}$, $\left( 1979 \right)$, pp.$1183-1201$.

\bibitem{hardywright} G.H.Hardy and E.M.Wright: {\it An introduction to the theory of numbers},
$5^{th}$ edition Oxford Univ. Press.

\bibitem{herman1} M.R.Herman: {\it Sur les courbes invariantes par les diff\'eomorphismes de l'anneau}, Ast\'erisque $\mathbf{103-104}$, $\left( 1983 \right)$, volume $1$.

\bibitem{Kolmogorov} A.N. Kolmogorov: {\it Preservation of conditionally periodic movements with
small change in the Hamilton function}, Dokl. Akad. Nauk SSSR, $\mathbf{98}$, $\left( 1954 \right)$,
pp.$527-530$.

\bibitem{Laskar1} J.Laskar: {\it Frequency analysis for multi-dimensional systems. Global dynamics
and diffusion}, Physica D $\mathbf{67}$, $\left( 1993 \right)$, pp.$257-281$.

\bibitem{Laskar2} J.Laskar: {\it The chaotic motion of the solar system. A numerical estimate of the
size of the chaotic zones}, Icarus $\mathbf{88}$, $\left( 1990 \right)$, pp.$266-291$.

\bibitem{Laskar3} J.Laskar: {\it Introduction to Frequency Map Analysis}, {\em In the proceedings of
3DHAM95 NATO Advanced Institute $533$ , S'Agaro, June 1995} $\left( 1999 \right)$ pp.$134-150$.

\bibitem{Laskar4} J.Laskar: {\it DA METTERE XXX}.

\bibitem{LaskarFroeschleCelletti} J.Laskar, C.Froeschl\'e and A.Celletti: {\it  The measure of chaos
by the numerical analysis of the fundamental frequencies. Applications to the standard mapping},
Physica D $\mathbf{56}$ $\left( 1992 \right)$, pp.$253-269$.



\bibitem{LaskarRobin} J.Laskar and D.Robin:{\it  Application of frequency map analysis to the ALS},
Particle Accelerator $\left( 1996 \right)$, $\mathbf{54}$, pp.$183-192$.

\bibitem{Lazutkin2} V.F.Lazutkin: {\it Splitting of separatrices for the Chirikov's standard map},
VINITI no. $6372/84$, $\left( 1984\right)$, (Russian).

\bibitem{Lazutkin} V.F.Lazutkin: {\it On the width of the instability zone near the separatrices of
a standard mapping}, Soviet. Math.Dokl. Vol. $\mathbf{42}$, $\left(1991\right)$, no. $1$.

\bibitem{Lazutkinschtab} V.F.Lazutkin,I.G.Schachmannski and M.B.Tabanov: {\it Splitting of
separatrices for standard ans semistandard mappings}, Physica D $\mathbf{40}$ $\left( 1989 \right)$
pp.$235-248$.

\bibitem{mackay} R.S. Mac Kay: {\it Exact results for an approximate renormalisation scheme and some predictions for the breackup of invariant tori}, Physica D $\mathbf{33}$ $\left( 1988 \right)$ pp. $240-265$. 

\bibitem{Marmi1} S.Marmi: {\it Critical functions for complex analytic maps}, J.Phys. A: Math. Gen.
$\mathbf{23}$ $\left( 1990 \right)$, pp.$3447-74$.

\bibitem{MarmiStark} S.Marmi J.Stark: {\it On the standard map critical function}, Nonlinearity
$\mathbf{5}$, $\left(1992\right)$ pp.$743-761$.

\bibitem{MMY} S.Marmi, P.Moussa and J.-C. Yoccoz: {\it The Brjuno functions and their regularity
properties}, Communications in Mathematical Physics $\mathbf{186}$, $(1997)$, pp.$265-293$.

\bibitem{Mather} J.N.Mather: {\it Non--existence of invariant circles}, Erg.Th. and Dyn.Syst.
$\mathbf{4}$, $(1984)$, pp.$301-309$.

\bibitem{Mather2} J.N.Mather: {\it A criterion for the non--existence of invariant circles}, Extrait
des Publications Math\'ematiques Insitut des Hautes \'Etudes Scientifiques, $\mathbf{63}$, $(1985)$.

\bibitem{Moser} J.Moser: {\it On invariant curves of area--preserving mapping of an annulus}, Nachr.
Akad. Wiss. G\"{o}tt,. II Math. Phys. Kl $1962$, $(1962)$, pp.$1-20$.

\bibitem{PAP96}
{Papaphilippou, Y., Laskar, J.}:  {\it Frequency map analysis and global dynamics
in a two degrees of freedom galactic potential},
{\it Astron. Astrophys.}, {\bf 307}, 427--449 (1996)

\bibitem{PAP98}
{Papaphilippou, Y., Laskar, J.}:  {\it Global dynamics of triaxial galactic models through
frequency map analysis},  {\it Astron. Astrophys.}, {\bf 329}, 451--481 (1998)


\bibitem{poincare1} H.Poincar\'e: {\it Les m\'ethodes nouvelles de la M\'echaniques C\'eleste}, vol
III (Paris: Gauthier Villars), $\left( 1892 \right)$.

\bibitem{poincare2} H.Poincar\'e: {\it Sur un Th\'eor\`eme de g\'eom\'etrie}, Rend.Circ.Mat.
Palermo, $\mathbf{33}$, $\left( 1912\right)$, pp.$375-407$.

\bibitem{poschel} J.P\"oschel: {\it Integrability of Hamiltonian systems on Cantor sets}, Comm.Pure
Appl.Math., $\mathbf{25}$, $(1982)$, pp. $653-695$.

\bibitem{TreschevZubelevich} D.Treschev and O.Zubelevich: {\it Invariant tori in Hamiltonian systems
with two degrees of freedom in a neighbourhood of a resonance}, Regular and chaotic dynamics, V.$3$,
N.$3$, $(1998)$.

\bibitem{yoccoz2} J.-C.Yoccoz: {\it Th\'eor\`eme de Siegel, polyn\^omes quadratiques et nombres de
Brjuno}, Ast\'erisque $\mathbf{231}$, $\left( 1995 \right)$, pp.$3 - 88$.

\end{thebibliography}
\end{document}